\documentclass[aps,pre,preprint,superscriptaddress]{revtex4-2}

\usepackage{amsmath,amssymb,amsthm}
\usepackage{graphicx}
\usepackage{bm}
\usepackage{hyperref}
\usepackage{color}

\begin{document}

\title{Identifiability of Contagion Components amid Environmental Fluctuations in Aggregated Default Counts}

\author{Shintaro Mori}
\affiliation{Graduate School of Science and Technology, Hirosaki University}

\begin{abstract}
Can contagion components be identified in aggregated default counts when default probabilities fluctuate with the economic environment? We study this question as an identifiability problem for coarse-grained default-count distributions. 
{
We adopt a binomial mixture with a Probit--Normal default probability as a parsimonious null model for environmental fluctuations and test whether contagion provides additional identifiable structure. Against this null, we consider two contagion mechanisms: cumulative contagion in the Davis--Lo model and threshold-type contagion in the Torri model. The Davis--Lo extension substantially improves the likelihood and information criteria for the ALL and speculative-grade classes, whereas the Torri component is largely absorbed by environmental heterogeneity; evidence is weaker for investment-grade defaults. As a robustness test, we replace the Gaussian environmental distribution by a two-component Gaussian mixture, and the Davis--Lo advantage persists for ALL and speculative-grade defaults. These results indicate that cumulative-contagion signatures remain distinguishable from the environmental null models considered here in the ALL and speculative-grade classes, even under annual aggregation.}
\end{abstract}

\keywords{default clustering, contagion, fluctuations, tail risk, credit risk}

\maketitle

\section{Introduction}
\label{sec:introduction}

In econophysics and sociophysics, macroscopic phenomena such as price fluctuations and collective behavior are often described as emergent consequences of interactions among heterogeneous agents \cite{MantegnaStanley1999,Galam2008,Lux1995,LuxMarchesi1999,Alfarano2005,Bouchaud2002,MoriHisakadoTakahashi2012,FernandezGracia2014}. However, similar aggregate patterns can arise from qualitatively different microscopic mechanisms. This makes the identification of the microscopic origin of observed fluctuations a central challenge, particularly when only aggregate observations are available \cite{SmolyakHavlin2022}.

This issue is particularly relevant in credit risk, where default clustering is widely observed
\cite{Schonbucher2003,DasDuffieKapadiaSaita2007}.
Firms are connected {
through business relationships and shared financial exposures, and defaults are therefore unlikely
to occur independently.}
Two broad classes of mechanisms have been proposed to explain such clustering: contagion, in which
the default of one firm directly affects others, and 
environmental fluctuations, {
in which common changes in the economic
environment simultaneously increase default probabilities across firms}~\cite{DavisLo2001,Vasicek1991,Vasicek2002,Vasicek2015}.
Firm-level and event-time approaches, including Hawkes-type self-exciting point processes, provide
important frameworks for tracing the timing and propagation of defaults when detailed micro-level
data are available
\cite{Hawkes1971,ErraisGieseckeGoldberg2010,Kirchner2017,BlancDonierBouchaud2017}.
{
More broadly, self-exciting mechanisms are widely studied in physical, societal, and biological
systems involving triggered events, including seismic activity, social interactions, epidemic
spreading, and financial events
\cite{KoyamaShinomoto2020,KanazawaSornette2020}.
Related empirical credit-risk studies have also examined whether observed default clustering can be
explained by observable or latent common factors, or whether an additional contagion component
is required
\cite{DuffieEcknerHorelSaita2009,AzizpourGieseckeSchwenkler2018}.
}

However, such data are often incomplete or unavailable over long historical periods. Many empirical datasets contain only aggregated default counts, with no information on inter-firm linkages or precise event timing.
{
Aggregation acts as a coarse-graining procedure that suppresses temporal and structural
information and reduces the observations to a default-count distribution. For exchangeable Bernoulli variables, de Finetti's theorem provides a conceptual basis for representing dependence through a latent-probability mixture. We use this observation only to motivate the mixture-distribution viewpoint; the contagion models considered below are not assumed to arise directly from this representation \cite{DeFinetti1974,HisakadoKitsukawaMori2006}. This mixture perspective also illustrates how distinct dependence mechanisms can generate very similar aggregate count distributions
\cite{HisakadoKitsukawaMori2006,Mori2008}, further limiting the extent to which microscopic
mechanisms can be distinguished from aggregated observations. Even if contagion exists at the micro level, it may not remain identifiable after aggregation.
In that setting, the relevant question is what can be identified about contagion and environmental fluctuations from coarse-grained observations.
}

{
In the present analysis, we deliberately do not impose a temporal-dynamics model on the annual latent environmental states. The annual sample contains only 104 observations, and introducing a specific dynamic law would add parameters and structural restrictions that may themselves be only weakly identified from such coarse and limited data and may make the resulting decomposition specification-dependent. Instead, we allow cross-year environmental heterogeneity to absorb aggregate variation without imposing a particular temporal dependence structure, and ask whether a residual contagion component remains distinguishable beyond this parsimonious environmental baseline.}

In this paper, we study this identifiability problem using historical annual default-count data.
We first compare the Davis--Lo cumulative-contagion model, the Torri threshold-type contagion
model, and the Vasicek common-factor model as static benchmarks. The Vasicek model provides
the best static fit. We then exploit an exact one-period equivalence: the Vasicek default-count
distribution can be represented as a binomial mixture whose default probability follows a
Probit--Normal distribution. This representation provides a parsimonious null model for
environmental fluctuations.

The central question is then whether cumulative or threshold-type contagion contributes
additional distributional structure beyond this environmental baseline. The cumulative-contagion
specification yields a substantial improvement for the ALL and speculative-grade classes,
whereas the threshold-type mechanism is more readily absorbed by environmental heterogeneity.
Evidence for investment-grade defaults is substantially weaker. The main conclusion remains
unchanged when the environmental null is replaced by a more flexible two-component Gaussian mixture.

This paper is organized as follows. Section II describes the data and presents summary statistics. Section III presents the three models and their static empirical
benchmark. Section IV examines identification relative to the Probit--Normal environmental null.
Section V tests robustness to a more flexible environmental null. Section VI discusses the
implications and concludes. Selected derivations and supplementary empirical results are collected
in the Appendix.

\section{Data description and summary statistics}
\label{sec:data}

\subsection{Data description}

We use annual default count data from Moody's covering the period 1920--2023.
{
The dataset covers corporate debt issuers rated by Moody's globally; it does not represent the universe of all firms. The aggregated annual data used here do not separately identify whether issuers are public or private, nor do they provide geographical information at the issuer level.
}
For each year $t$, the dataset provides the total number of obligors $n_t$ and the number of defaults $L_t$ for speculative-grade (SG), investment-grade (IG), and aggregate (ALL) categories. The default rate is defined as $L_t / n_t$.

Since only aggregated default counts are available, the underlying network structure and the timing of individual defaults are not observed. This limitation motivates our focus on distributional properties and identification based on count data.

\subsection{Summary statistics}

Table~\ref{tab:summary_period} reports summary statistics for the full sample
(1920--2023) and for two selected subperiods (1950--1979 and 1980--2023).
{
The two subperiods provide a historically motivated comparison outside the exceptional period spanning the Great Depression and World War II. The year 1980 is used as a convenient boundary between a period of relatively low default activity and the subsequent period of higher and more variable default activity; it is not intended as an estimated structural break.}
The mean default rate is the simple average of yearly default rates, whereas
the total default rate is the exposure-weighted average over the corresponding period.

\begin{table}[htbp]
\centering
\caption{Summary statistics of annual default counts for the full sample and selected subperiods.}
\label{tab:summary_period}
\begin{tabular}{llcccc}
\hline
Class & Period & Mean $n$ & Mean $L$ & Mean default rate & Total default rate \\
\hline
ALL & 1920--2023 & 2573 & 40.30 & 0.0120 & 0.0157 \\
    & 1950--1979 & 860.2 & 2.53 & 0.0026 & 0.0029 \\
    & 1980--2023 & 4276.4 & 71.86 & 0.0157 & 0.0168 \\
\hline
SG  & 1920--2023 & 1008 & 38.13 & 0.0288 & 0.0378 \\
    & 1950--1979 & 240.0 & 2.37 & 0.0093 & 0.0099 \\
    & 1980--2023 & 1678.4 & 69.32 & 0.0416 & 0.0413 \\
\hline
IG  & 1920--2023 & 1565 & 2.17 & 0.0014 & 0.0014 \\
    & 1950--1979 & 620.2 & 0.17 & 0.0002 & 0.0003 \\
    & 1980--2023 & 2598.0 & 2.55 & 0.0009 & 0.0010 \\
\hline
\end{tabular}
\end{table}

The subperiod comparison reveals substantial temporal variation.
In particular, the mean SG default rate increases from 0.0093 in 1950--1979
to 0.0416 in 1980--2023, whereas the IG default rate remains much lower,
increasing from 0.0002 to 0.0009.

\subsection{Time series properties}

\begin{figure}[htbp]
\centering
\includegraphics[width=0.85\textwidth]{fig_default_rate_timeseries.pdf}
\caption{
Time series of annual default rates for ALL, SG, and IG portfolios.
The shaded region corresponds to 1950--1979.
Default clustering is particularly evident for SG, while IG exhibits low default rates with occasional spikes.
}
\label{fig:default_ts}
\end{figure}

Figure~\ref{fig:default_ts} shows the time series of default rates for ALL, SG, and IG portfolios. The shaded region corresponds to the early subperiod (1950--1979).

The series exhibit pronounced clustering of defaults over time, with periods of elevated default activity followed by relatively calm phases. This clustering is particularly evident for SG portfolios, while IG portfolios display low default rates with occasional spikes.
These features motivate the subperiod robustness analysis presented later.

\section{Models and Static Benchmark}
\label{sec:models}

We consider { three one-period models for the number of defaults in a homogeneous portfolio of $n$ obligors:} the Infectious Default Model of Davis and Lo (hereafter, the Davis--Lo model) \cite{DavisLo2001}, the Infectious Default Model with Immunization of Torri, Giacometti, and Farina (hereafter, the Torri model) \cite{TorriGiacomettiFarina2026}, and the Vasicek common-factor model \cite{Vasicek1991}. Throughout this paper, we consider homogeneous portfolios, in which obligors share the same baseline default probability and the same model-specific dependence parameters. This assumption allows us to focus on the identifiability of dependence mechanisms in aggregated default-count distributions.

The three models generate dependence in different ways. In the Davis--Lo model, contagion is cumulative: the probability of additional defaults increases with the number of seed defaults. In the Torri model, contagion is activated by the existence of at least one infectious default, producing a threshold-type mechanism. In the Vasicek model, defaults are conditionally independent given a continuous common factor, which generates a {continuous} binomial mixture.

{
Detailed expressions for the moment generating functions, mean default probabilities, and pairwise dependence are given in Appendix~\ref{app:models}.
}
We first compare these three one-period models as static benchmarks. The environmental-null framework used for the identifiability analysis is introduced in Sec.~\ref{sec:identification}.

\subsection{Davis--Lo Cumulative Contagion}
\label{subsec:davis-lo}
For each obligor $i = 1, \dots, n$, let $X_i \in \{0,1\}$ denote the idiosyncratic default indicator, with $X_i \sim \mathrm{Bernoulli}(p)$. For each ordered pair $(i,j)$ with $i \ne j$, let $Y_{ij} \in \{0,1\}$ denote the contagion indicator from obligor $j$ to obligor $i$, with $Y_{ij} \sim \mathrm{Bernoulli}(q)$. All variables are assumed to be independent.

The final default indicator is defined as
\begin{equation}
Z_i = X_i + (1 - X_i)\left(1 - \prod_{j \ne i}(1 - Y_{ij} X_j)\right).
\end{equation}

Let $L_n = \sum_{i=1}^n Z_i$ denote the total number of defaults, and let $K = \sum_{i=1}^n X_i$ denote the number of idiosyncratic defaults. 
Then, by independence, $K \sim \mathrm{Binomial}(n,p)$.
Conditional on $K = k$, each surviving obligor defaults independently with probability
\begin{equation}
r_k = 1 - (1 - q)^k,
\end{equation}
so that
\begin{equation}
L_n \mid K = k \sim k + \mathrm{Binomial}(n-k, r_k).
\end{equation}
This representation highlights the cumulative nature of contagion: the probability of additional defaults $r_k$ increases smoothly with the number of {seed defaults} $k$. { The independent-binomial model is recovered at $q=0$.}

\subsection{Torri Threshold-Type Contagion}
\label{subsec:torri}
For each obligor $i = 1, \dots, n$, let $X_i \in \{0,1\}$ denote the idiosyncratic default indicator, with $X_i \sim \mathrm{Bernoulli}(p)$. In addition, let $U_i \in \{0,1\}$ denote the immunization indicator, with $U_i \sim \mathrm{Bernoulli}(u)$, where $U_i = 1$ indicates that obligor $i$ is immune to contagion. Finally, let $V_i \in \{0,1\}$ denote the infectivity indicator, with $V_i \sim \mathrm{Bernoulli}(v)$, meaning that obligor $i$ can act as a contagion trigger when it defaults idiosyncratically. All variables are assumed to be independent.

The final default indicator is defined as
\begin{equation}
Z_i = X_i + [1 - X_i] \cdot [1 - U_i] \cdot 
\left\{1 - \prod_{j \ne i} \left[1 - X_j V_j\right]\right\}.
\end{equation}

This expression shows that contagion occurs if there exists at least one obligor $j$ such that $X_j=1$ and $V_j=1$. 
Accordingly, we define the global contagion indicator
\[
I_n^C = \mathbf{1}\!\left(\sum_{j=1}^n X_jV_j > 0\right),
\]
and the model can be equivalently written as
\begin{equation}
Z_i = X_i + [1-X_i] [1-U_i] I_n^C.
\end{equation}
{
The activation probability of the contagious state is therefore}
\begin{equation}
\pi_n := \mathbb{P}(I_n^C=1) = 1-(1-pv)^n.
\end{equation}

{
Conditional on $K=k$ idiosyncratic defaults, contagion is inactive with probability $(1-v)^k$. If it is active, each of the $n-k$ surviving obligors defaults with probability $1-u$. Therefore,
\begin{equation}
P_{\mathrm{T}}(L_n=h\mid K=k)={}
(1-v)^k\mathbf{1}_{\{h=k\}}+
\left[1-(1-v)^k\right]
\binom{n-k}{h-k}
(1-u)^{h-k}u^{n-h}.
\label{eq:torri_conditional}
\end{equation}
Combining Eq.~(\ref{eq:torri_conditional}) with $K\sim\mathrm{Binomial}(n,p)$ gives the exact one-period default-count likelihood.

Unlike the Davis--Lo mechanism, the Torri mechanism is threshold-like: once at least one infectious seed activates the contagious state, additional infectious defaults do not strengthen contagion further. The no-contagion submodel is obtained at $v=0$ or $u=1$. In particular, when $v=0$, contagion is never activated and $u$ is not identified. }

\subsection{Vasicek Model and Probit--Normal Representation}
\label{subsec:vasicek}
The Vasicek model generates dependence through a continuous latent factor. In contrast to contagion-based models, defaults are conditionally independent given a common macroeconomic factor. For each obligor $i = 1, \dots, n$, define the latent variable
\begin{equation}
Y_i = \sqrt{\rho_A}\,F + \sqrt{1-\rho_A}\,\varepsilon_i,
\end{equation}
where $F \sim \mathcal{N}(0,1)$ is a common factor and $\varepsilon_i \sim \mathcal{N}(0,1)$ are i.i.d. idiosyncratic shocks, independent of $F$. 
The parameter $\rho_A \in [0,1)$ represents the asset correlation. 
Let $\Phi$ denote the cumulative distribution function of the standard normal distribution. 
Obligor $i$ defaults when $Y_i \le \Phi^{-1}(p)$.
The corresponding default indicator is defined as
\[
Z_i = \mathbf{1}\{Y_i \le \Phi^{-1}(p)\},
\]
so that $\mathbb{P}(Z_i=1)=p$.

Let $L_n = \sum_{i=1}^n Z_i$ denote the total number of defaults. Conditional on $F=f$, defaults are independent with probability
\begin{equation}
p(f) = \Phi\!\left(\frac{\Phi^{-1}(p) - \sqrt{\rho_A}\,f}{\sqrt{1-\rho_A}}\right),
\label{eq:vasicek_conditional_p}
\end{equation}
so that
\begin{equation}
L_n \mid F=f \sim \mathrm{Binomial}(n, p(f)).
\end{equation}
The unconditional distribution is therefore a continuous binomial mixture induced by the latent factor $F$. 

{
This representation has an exact reformulation that is central to the analysis below. Define
\begin{equation}
y=\frac{\Phi^{-1}(p)-\sqrt{\rho_A}F}
{\sqrt{1-\rho_A}}.
\end{equation}
Because $F\sim N(0,1)$,
\begin{equation}
y\sim N(\mu,\sigma^2),
\end{equation}
with
\begin{equation}
\mu
=
\frac{\Phi^{-1}(p)}
{\sqrt{1-\rho_A}},
\qquad
\sigma
=
\sqrt{\frac{\rho_A}{1-\rho_A}}.
\label{eq:vasicek_to_pn}
\end{equation}
Equations~(\ref{eq:vasicek_conditional_p})--(\ref{eq:vasicek_to_pn}) imply the equivalent representation
\begin{equation}
\tilde p=\Phi(y),
\qquad
L_n\mid \tilde p\sim\mathrm{Binomial}(n,\tilde p).
\label{eq:probit_normal_representation}
\end{equation}

Conversely,
\begin{equation}
p=\Phi\left(\frac{\mu}{\sqrt{1+\sigma^2}}\right),
\qquad \rho_A=
\frac{\sigma^2}{1+\sigma^2}.
\label{eq:pn_to_vasicek}
\end{equation}

Thus the Vasicek common-factor model and the Probit--Normal binomial mixture generate exactly the same one-period default-count likelihood. This equivalence does not imply the same microscopic interpretation. Rather, it allows the empirically successful Vasicek count distribution to be reinterpreted as a parsimonious model of environmental variation in the default probability. This observation provides the null model used in Section~\ref{sec:identification}.
}

Additional comparisons of the one-period model families based on directed
Kullback--Leibler (KL) divergence are reported in Appendix~\ref{app:models}
as descriptive diagnostics of their distributional relations.

\subsection{Empirical Fit in the Static Setting}
\label{sec:iid-fit}

Using the data described in Sec.~\ref{sec:data}, we first fit the three one-period models in the static setting.
{
The $T=104$ yearly observations are treated as independent conditional on the observed portfolio sizes $n_t$, with common model parameters across years.
}
This analysis provides a benchmark for assessing which aspects of the empirical distribution can be captured without explicitly modeling time variation.

Model parameters are estimated by maximum likelihood:
\begin{equation}
\ell(\theta)=
\sum_{t=1}^{T}
\log P(L_t\mid n_t,\theta).
\end{equation}

Table~\ref{tab:iid_fit_moments} reports the maximum-likelihood estimates,
negative log-likelihoods, and implied moments in the static setting.
Across all classes, the Vasicek model achieves the lowest negative
log-likelihood, followed by the Davis--Lo model, while the Torri model gives the
poorest fit. The empirical rows provide the corresponding mean default
probabilities and pairwise default correlations computed from the annual count
data.

\begin{table}[htbp]
\centering
\caption{
Maximum-likelihood estimates and implied moments in the static setting.
For each class, the lowest negative log-likelihood is shown in bold.
The quantities $m$ and $\rho$ denote the mean default probability and pairwise
default correlation, respectively. For the data rows, $m$ and $\rho$ are computed
from the empirical annual default counts using the same aggregation scheme as in
Appendix~\ref{app:add_iid}.
}
\label{tab:iid_fit_moments}
\begin{tabular}{llccccc}
\hline
Class & Source & nll & $p$ & Dep. param. & $m$ & $\rho$ \\
\hline
ALL & Data      & -- & -- & -- & 0.0157 & 0.01134 \\
ALL & Davis--Lo & 461.2 & 0.00113 & $q=0.00379$ & 0.0183 & 0.01605 \\
ALL & Torri     & 725.8 & 0.00840 & $u=0.98,\ v=0.014$ & 0.0157 & 0.00958 \\
ALL & Vasicek   & \textbf{431.4} & 0.0118 & $\rho_A=0.136$ & 0.0118 & 0.01542 \\
\hline
IG & Data      & -- & -- & -- & 0.0014 & 0.00191 \\
IG & Davis--Lo & 194.5 & 0.00046 & $q=0.00101$ & 0.0016 & 0.00201 \\
IG & Torri     & 290.3 & 0.00139 & $u=0.58,\ v\simeq 0$ & 0.0014 & $3.947\times 10^{-11}$ \\
IG & Vasicek   & \textbf{182.4} & 0.00161 & $\rho_A=0.200$ & 0.0016 & 0.00798 \\
\hline
SG & Data      & -- & -- & -- & 0.0378 & 0.02202 \\
SG & Davis--Lo & 442.0 & 0.00280 & $q=0.00924$ & 0.0476 & 0.05045 \\
SG & Torri     & 643.3 & 0.0223 & $u=0.95,\ v=0.012$ & 0.0387 & 0.02363 \\
SG & Vasicek   & \textbf{417.7} & 0.0304 & $\rho_A=0.150$ & 0.0304 & 0.03089 \\
\hline
\end{tabular}
\end{table}

To compare model-implied distributions with empirical data while controlling for variation in portfolio size, we fix the portfolio size at the sample mean
\[
\bar n = \frac{1}{T} \sum_{t=1}^T n_t,
\]
and rescale the observed counts as
\[
L_t^{(\mathrm{scaled})}
= \mathrm{round}\!\left(
L_t \frac{\bar n}{n_t}
\right)
\]
where $\operatorname{round}(x)=\lfloor x+1/2\rfloor$ for $x\ge0$.

\begin{figure}[htbp]
\centering
\includegraphics[width=0.9\linewidth]{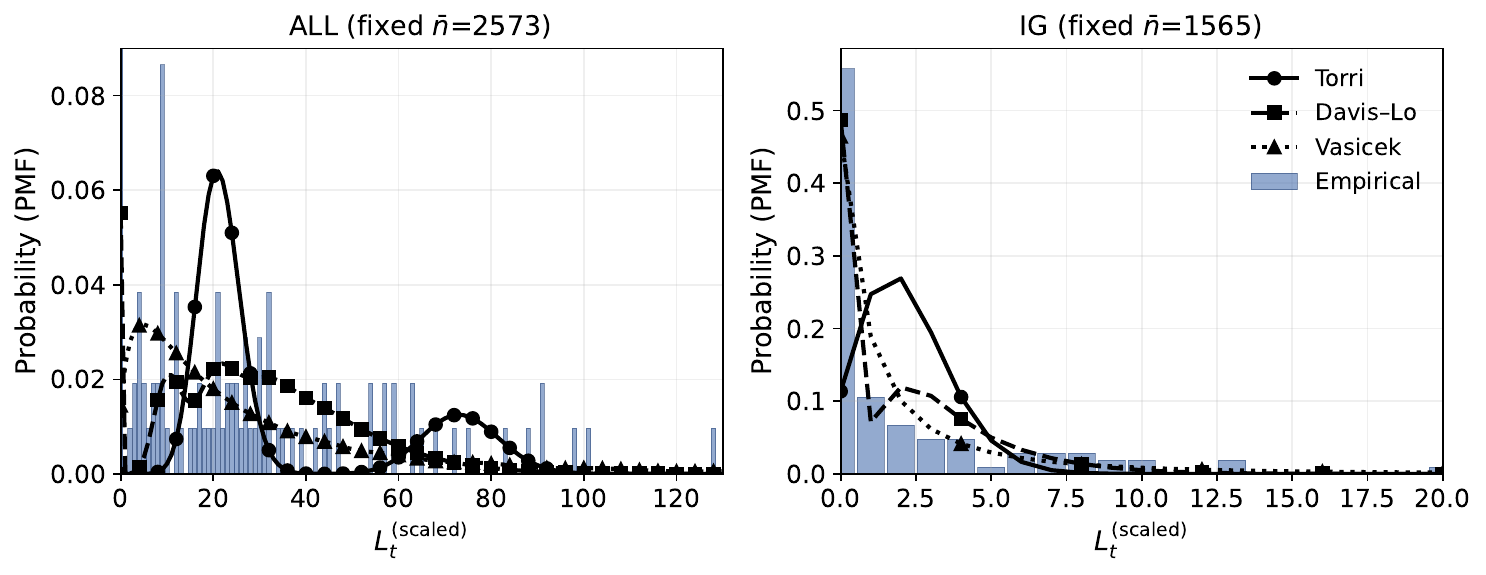}

\includegraphics[width=0.9\linewidth]{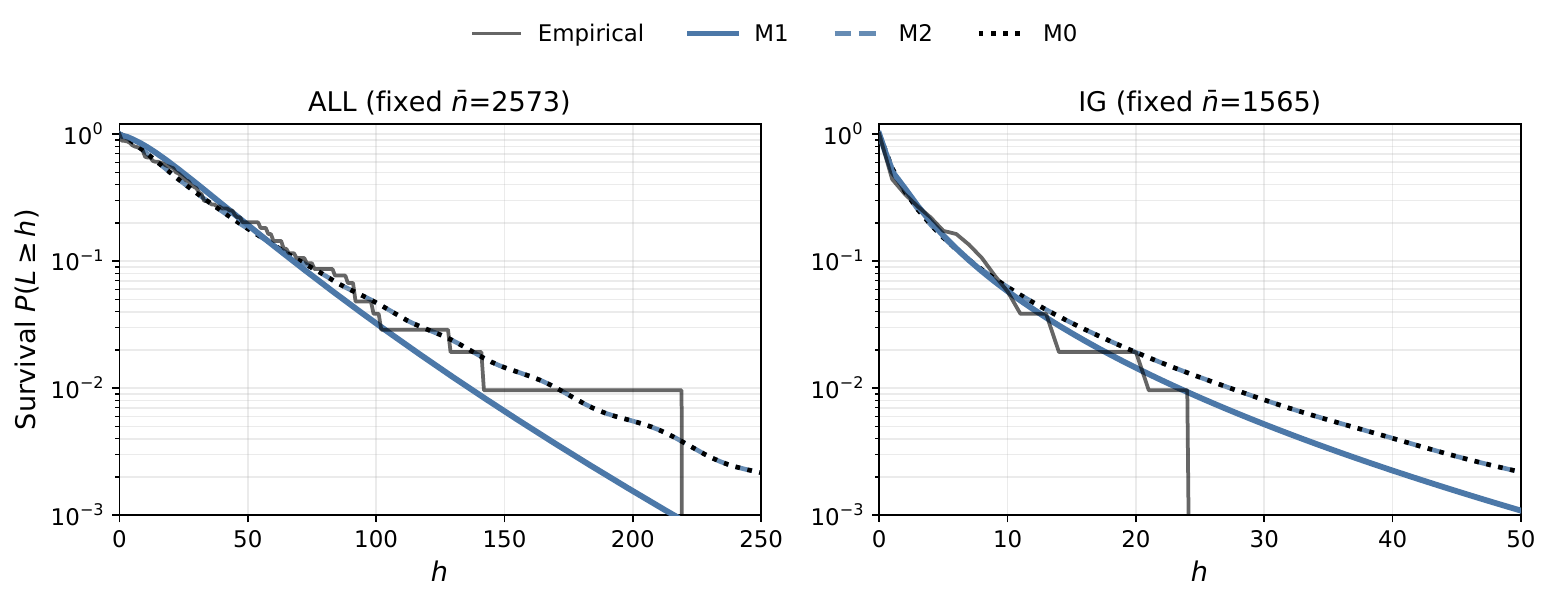}

\caption{
Comparison of model-implied distributions and tail behavior at the fixed portfolio size $n=\bar n$ for the ALL and IG classes.
\textbf{Top:} Probability mass functions (PMFs).
\textbf{Bottom:} Survival functions $S(h)=P(L \ge h)$ on a logarithmic scale.
{
The Torri model places relatively high probability in an intermediate range of default counts and then exhibits a rapidly decaying tail, thereby underestimating large default events.
The Davis--Lo model produces a smoother tail but still underestimates sufficiently extreme events, whereas the Vasicek model provides the closest overall description of the empirical tail behavior.
}
}
\label{fig:pmf_survival_all_ig}
\end{figure}

Figure~\ref{fig:pmf_survival_all_ig} shows the comparison between model-implied distributions and the rescaled data.
{
The models differ most clearly away from the center of the distribution.
The Torri model assigns relatively high probability to an intermediate range of default counts but then exhibits a rapid decay of tail probabilities, leading to underestimation of large default events.
The Davis--Lo model produces a smoother distribution, but its tail also becomes too light for sufficiently extreme counts.
In contrast, the Vasicek model generates a more gradual tail decay and provides the closest overall agreement with the empirical survival function.
}

{
This difference is particularly visible in the IG class, where the distribution is concentrated at low default counts but contains occasional large-default years.
}
Overall, these comparisons show that, in the static setting,
{the Vasicek continuous-mixture model} provides the closest representation of the empirical default-count distribution.

{
The static superiority of the Vasicek model should not, however, be interpreted as evidence that a Vasicek common factor is the true microscopic source of default clustering. At the level of one-period aggregated counts, the model is exactly equivalent to the Probit--Normal environmental mixture derived above. The static benchmark therefore motivates a different identification question: after adopting this empirically successful environmental mixture as the null model, is there still evidence for an additional contagion component? Section~\ref{sec:identification} addresses this question directly.}


\section{Identification Relative to the Probit--Normal Environmental Null}
\label{sec:identification}

The preceding section examined the empirical performance of the Davis--Lo,
Torri, and Vasicek models in the static setting.
The Vasicek model provided the best overall fit, {suggesting that}
a smooth mixture mechanism can absorb a substantial part of the variation in
aggregated annual default counts.

We now introduce cross-period variation in the baseline default probability
explicitly.
The {M0--M2 framework} considered below superposes environmental
fluctuations of the baseline default probability on the Davis--Lo and Torri
default-count distributions.
The question in this section is whether any residual Davis--Lo or Torri
contagion component remains distinguishable after this environmental mixture
component is added.

\subsection{Environmental Null and Contagion Extensions}

In empirical annual data, default conditions vary across years.
To account for this effect, we {allow the baseline default probability
to fluctuate across periods while the structural contagion parameters, when
present, are kept fixed.}

For each period $t$, we write
\begin{equation}
y_t \sim N(\mu,\sigma^2),
\qquad
p_t=\Phi(y_t).
\label{eq:pn_environment}
\end{equation}

{
The index $t$ labels annual observations only: no temporal dynamics are
introduced, and the latent environmental states are treated as independent
across years.
The Probit--Normal environmental null, denoted by M0, is
\begin{equation}
L_t\mid p_t,n_t
\sim
\mathrm{Binomial}(n_t,p_t).
\label{eq:m0_conditional}
\end{equation}
By the equivalence established in Sec.~\ref{subsec:vasicek}, M0 generates
exactly the same one-period default-count likelihood as the Vasicek model,
with two free parameters $(\mu,\sigma)$.
}

{
We consider two extensions of this null.
Model M1 adds Davis--Lo cumulative contagion: conditional on $p_t$, the
one-period Davis--Lo distribution of Sec.~\ref{subsec:davis-lo} is used with
$p=p_t$ and a time-independent contagion parameter $q$.
Model M2 instead adds Torri threshold-type contagion, using $p=p_t$ together
with time-independent parameters $(u,v)$.
Thus,
\begin{equation}
P_j(L_t=h\mid n_t)
=\int_{-\infty}^{\infty}
P_j\!\left(h\mid n_t,\Phi(y),\eta_j\right)
\frac{1}{\sigma}
\phi\!\left(\frac{y-\mu}{\sigma}\right)\,dy,
\label{eq:pn_contagion_mixture}
\end{equation}
where $\phi$ is the standard normal density and
\begin{equation}
\eta_0=\varnothing,
\qquad
\eta_1=q,
\qquad
\eta_2=(u,v).
\end{equation}
The parameter counts are therefore $k=2$, $3$, and $4$ for M0, M1, and M2,
respectively.
}

{For M1 and M2, }this construction superposes environment-induced
fluctuations of the baseline default probability on contagion-induced
default-count distributions.
This specification separates two sources of aggregate fluctuations:
within-period dependence generated by contagion and cross-period variation
in the baseline default probability.

{
Both contagion specifications contain the environmental null as a boundary
case.
For M1, $q=0$ removes contagion and reduces the model exactly to M0.
For M2, contagion is absent at $v=0$ or $u=1$; when $v=0$, the immunization
parameter $u$ no longer affects the likelihood and is unidentified.
This boundary structure is important for interpreting the empirical fits.
}

\subsection{Likelihood Comparison}

{
For each model, the log-likelihood is
\begin{equation}
\ell_j(\theta_j)
=\sum_{t=1}^{T}
\log P_j(L_t\mid n_t,\theta_j),
\qquad
T=104.
\label{eq:m012_loglik}
\end{equation}
We compare the maximized likelihoods using the Akaike and Bayesian information
criteria,
\begin{equation}
\mathrm{AIC}=2k-2\ell_{\max},
\qquad
\mathrm{BIC}=k\log T-2\ell_{\max}.
\label{eq:aic_bic}
\end{equation}
Because the no-contagion null lies on the boundary of the extended model
families, and because $u$ is unidentified for M2 when $v=0$, we do not rely
on a regular chi-square likelihood-ratio approximation.
Instead, the likelihood improvement and the information criteria are used
together to assess whether the additional contagion parameters are supported
by the annual count data.
}

{
Table~\ref{tab:m012_comparison} compares the empirical fit of the
Probit--Normal environmental null M0 with its Davis--Lo and Torri contagion
extensions, M1 and M2.
The corresponding maximum-likelihood estimates of the environmental
parameters $(\mu,\sigma)$ under M0 are
$(-2.435509,\,0.396041)$ for ALL,
$(-2.033720,\,0.419413)$ for SG, and
$(-3.292821,\,0.500001)$ for IG.
}

\begin{table}[t]
\centering
\caption{{
Comparison of the Probit--Normal environmental null (M0) with the
Davis--Lo (M1) and Torri (M2) contagion extensions.
Lower values of NLL, AIC, and BIC indicate better fit.
Bold entries denote the best AIC or BIC within each credit class.}
}
\label{tab:m012_comparison}
\begin{tabular}{llrrrr}
\hline
Class & Model & $k$ & NLL & AIC & BIC \\
\hline
ALL & {M0} & 2 & 431.364 & 866.727 & 872.016 \\
    & {M1} & 3 & 413.825 & \textbf{833.651} & \textbf{841.584} \\
    & {M2} & 4 & 431.364 & 870.727 & 881.305 \\
\hline
SG  & {M0} & 2 & 417.709 & 839.417 & 844.706 \\
    & {M1} & 3 & 401.342 & \textbf{808.684} & \textbf{816.618} \\
    & {M2} & 4 & 409.267 & 826.533 & 837.111 \\
\hline
IG  & {M0} & 2 & 182.382 & 368.763 & \textbf{374.052} \\
    & {M1} & 3 & 180.781 & \textbf{367.563} & 375.496 \\
    & {M2} & 4 & 182.382 & 372.763 & 383.341 \\
\hline
\end{tabular}
\end{table}

The {M1} model achieves the lowest negative log-likelihood in all
classes, with substantial gains for ALL and SG and a marginal gain for IG.
{
For ALL, M1 lowers the negative log-likelihood from $431.364$ to $413.825$,
and the improvement remains substantial after penalizing the additional
parameter in both AIC and BIC.
By contrast, the best-found M2 solution lies on the no-contagion boundary
and reproduces the M0 likelihood.
}

{
For SG, M1 again produces a substantial improvement over M0.
M2 also gives a non-trivial fit and improves on M0, indicating that
threshold-type contagion is not completely absorbed by environmental
heterogeneity in this class.
Nevertheless, M2 remains clearly inferior to M1 in likelihood, AIC, and BIC.
}

For the IG class, the negative log-likelihood of {M2} is almost
identical to that of {M0}.
This occurs because the Torri infectivity parameter collapses to the boundary
value $v=0$, so that the threshold-contagion mechanism is effectively switched
off.
Once $v=0$, the immunization parameter $u$ no longer affects the likelihood
and becomes unidentified.
The {M2 specification} therefore reduces to a smooth Probit--Normal
mixture of binomial distributions in this class.
{
M1 lowers the negative log-likelihood by only about $1.60$ relative to M0.
AIC slightly favors M1, whereas BIC favors the simpler M0 specification.
Hence the annual IG counts provide substantially weaker evidence for an
additional contagion component.
}

{
Taken together, the comparison is both mechanism- and class-dependent.
For ALL and SG, cumulative contagion produces a substantial residual
likelihood improvement after environmental heterogeneity is introduced.
Threshold-type contagion is more readily absorbed by the environmental
mixture: M2 collapses to M0 for ALL and IG and, although non-trivial for SG,
remains clearly inferior to M1.
The identification claim is therefore relative to the Probit--Normal
environmental null and to the model families considered, rather than a claim
to recover a unique microscopic mechanism.
}

\subsection{Tail Behavior and Distributional Implications}

We examine the tail behavior {of M0, M1, and M2} by comparing the
survival function of the scaled default counts together with the corresponding
Value-at-Risk (VaR) and Expected Shortfall (ES).
{
The same fixed-portfolio-size scaling as in Sec.~\ref{sec:iid-fit} is used.
}

\begin{table}[htbp]
\centering
\caption{
Tail-risk comparison {for the Probit--Normal environmental null
(M0), the Davis--Lo extension (M1), and the Torri extension (M2)}.
Bold model values indicate the closest value to the empirical value for each
metric.
}
\label{tab:tail_compare_m012}
\begin{tabular}{llcccccccc}
\hline
Class & Metric
& \multicolumn{2}{c}{Data}
& \multicolumn{2}{c}{{M1}}
& \multicolumn{2}{c}{{M2}}
& \multicolumn{2}{c}{{M0}} \\
\cline{3-4}\cline{5-6}\cline{7-8}\cline{9-10}
& $\alpha$ & VaR & ES & VaR & ES & VaR & ES & VaR & ES \\
\hline
ALL & 0.95 & 91  & 124.1 & \textbf{87}  & \textbf{117.4} & 97  & 141.4 & 97  & 141.4 \\
    & 0.99 & 141 & 180.0 & \textbf{136} & \textbf{169.7} & 170 & 220.9 & 170 & 220.9 \\
\hline
SG  & 0.95 & 92  & 113.7 & 81  & \textbf{106.1} & 113 & 159.3 & \textbf{91} & 125.5 \\
    & 0.99 & 122 & 140.5 & \textbf{123} & \textbf{149.2} & 190 & 239.2 & 148 & 184.3 \\
\hline
IG  & 0.95 & 10  & 15.0 & \textbf{10} & \textbf{17.4} & 11 & 20.7 & 11 & 20.7 \\
    & 0.99 & 20  & 22.0 & \textbf{23} & \textbf{34.2} & 27 & 42.1 & 27 & 42.1 \\
\hline
\end{tabular}
\end{table}

{
Table~\ref{tab:tail_compare_m012} shows that the gain from M1 is not confined
to the center of the distribution.
}
For the ALL and SG classes, the {M1} model generally provides the
closest agreement with the empirical VaR and ES, although the
{M0} model performs comparably at the 95\% VaR level.
{ For ALL, M2 collapses to M0 and therefore yields the same VaR and ES.
For SG, M2 remains distinct from M0 and produces substantially larger tail-risk
measures.}
In the IG class, {M2 also collapses to M0}, and the far-tail fit
remains difficult for all specifications, although the {M1}
specification is the least excessive.

\begin{figure}[t]
\centering
\includegraphics[width=0.9\linewidth]{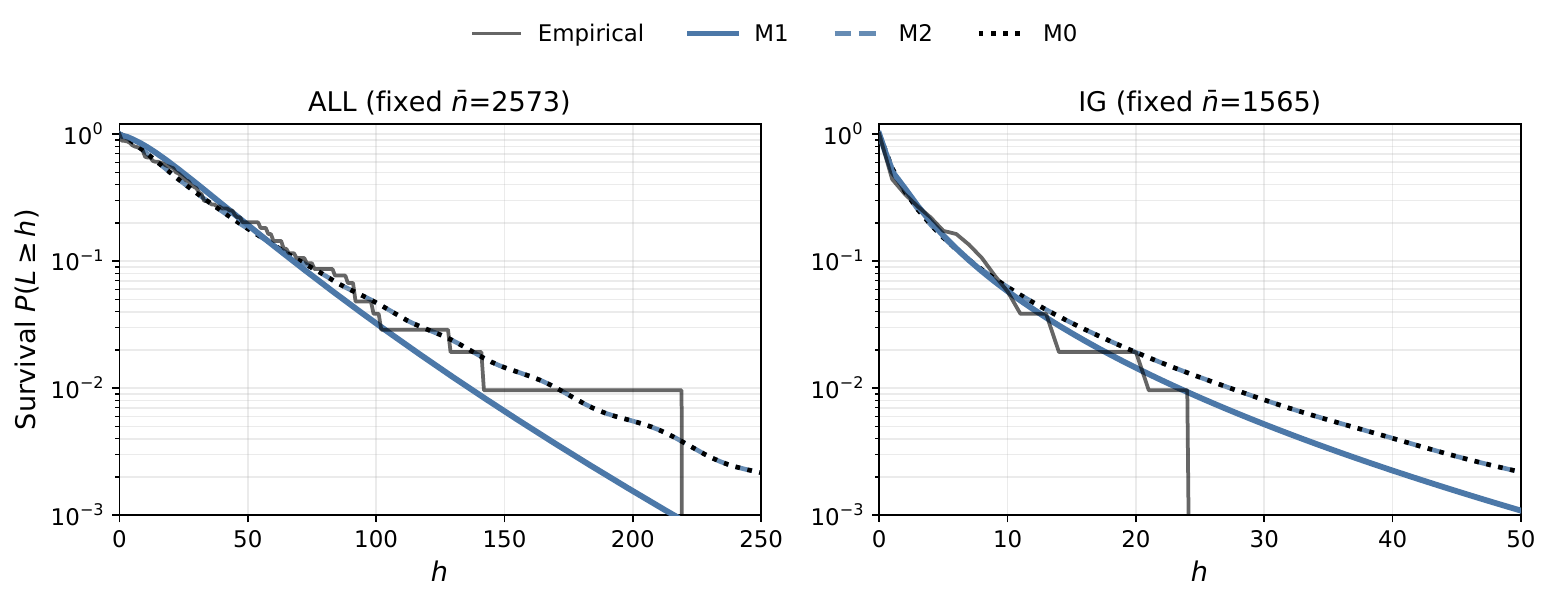}
\caption{
Survival functions of scaled default counts
{for the environmental null and its contagion extensions}.
The {M1} model closely matches the empirical tail behavior in the
ALL class, while {M2 coincides with M0}.
In the IG class, {M2 likewise coincides with M0}, and all
specifications tend to overestimate the far tail.
}
\label{fig:survival_m012}
\end{figure}

Figure~\ref{fig:survival_m012} confirms these findings.
In the ALL class, the {M1} model closely tracks the empirical
survival function over the observed range, 
whereas {M2 and M0 are distributionally identical}.
{Both} exhibit a slower decay than observed in
the data.
In the IG class, {M2 again coincides with M0}, while all models tend
to display overly heavy far tails, reflecting the sensitivity of rare-default
regimes to fluctuations in $p_t$.

{
Thus, for ALL and SG, the likelihood improvement of M1 over the environmental
null M0 has a visible distributional counterpart in the tail rather than
arising only from small adjustments near the center of the default-count
distribution.
}

{
Additional diagnostics based on variance decomposition and directed Kullback--Leibler (KL) divergence among the fitted M0--M2 distributions are reported in Appendix~\ref{app:hier_additional}. The variance decomposition shows that fluctuations in the annual default probability provide the largest normalized contribution to the empirical variance, whereas the residual contagion contribution is substantially smaller and mechanism-dependent.
}

{
The remaining concern is whether this result depends on the single-Gaussian
assumption for the latent environmental state in
Eq.~\eqref{eq:pn_environment}.
Section~\ref{sec:robustness} addresses this point by replacing the Gaussian
environmental null with a more flexible two-component Gaussian mixture.
}


\section{Robustness to a Flexible Environmental Null}
\label{sec:robustness}
{

The identification results in Sec.~\ref{sec:identification} are relative to a
Probit--Normal environmental baseline. A natural concern is that the residual
Davis--Lo improvement may arise because the single-Gaussian distribution of the
latent environmental state is too restrictive. We therefore perform a direct
robustness check by increasing the flexibility of the environmental null while
leaving out contagion entirely.

\subsection{Two-component Gaussian-mixture environmental null}

Model M3 replaces the single Gaussian distribution in M0 by a two-component
Gaussian mixture on the probit scale,
\begin{equation}
y_t \sim
\pi N(\mu_1,\sigma_1^2)
+(1-\pi)N(\mu_2,\sigma_2^2),
\qquad
p_t=\Phi(y_t),
\label{eq:m3_environment}
\end{equation}
with
\begin{equation}
L_t\mid p_t,n_t\sim \mathrm{Binomial}(n_t,p_t).
\label{eq:m3_conditional}
\end{equation}
The model has five free parameters,
$(\mu_1,\sigma_1,\mu_2,\sigma_2,\pi)$, with $0<\pi<1$ and
$\sigma_1,\sigma_2>0$. For reporting we order the component means as
$\mu_1<\mu_2$. The M0 family is recovered in the closure of M3 when the two
Gaussian components coincide, so M3 provides a direct test of whether additional
environmental flexibility can absorb the residual distributional structure
attributed to contagion in M1.

For an annual observation, the M3 count likelihood is
\begin{align}
P_3(L_t=h\mid n_t)
={}&\pi\int
\mathrm{Bin}\!\left(h\mid n_t,\Phi(y)\right)
\varphi(y;\mu_1,\sigma_1^2)\,dy
\nonumber\\
&+(1-\pi)\int
\mathrm{Bin}\!\left(h\mid n_t,\Phi(y)\right)
\varphi(y;\mu_2,\sigma_2^2)\,dy,
\label{eq:m3_likelihood}
\end{align}
where $\varphi(\cdot;\mu,\sigma^2)$ denotes a normal density. As in the previous
sections, the likelihood uses the observed yearly portfolio sizes $n_t$.

\subsection{Model comparison}

Table~\ref{tab:m3_comparison} compares the flexible environmental null M3 with
M0 and with the Davis--Lo extension M1. The M3 values are cross-language
validated best-found likelihood candidates. Because finite-mixture likelihoods
can contain multiple nearby modes, they are not described as guaranteed global
optima.

\begin{table}[t]
\centering
\caption{
Comparison of the single-Gaussian environmental null (M0), the Davis--Lo
extension (M1), and the two-component Gaussian-mixture environmental null (M3).
Lower NLL, AIC, and BIC indicate better fit.
}
\label{tab:m3_comparison}
\begin{tabular}{llrrrr}
\hline
Class & Model & $k$ & NLL & AIC & BIC \\
\hline
ALL & M0 & 2 & 431.364 & 866.727 & 872.016 \\
    & M1 & 3 & \textbf{413.825} & \textbf{833.651} & \textbf{841.584} \\
    & M3 & 5 & 430.271 & 870.543 & 883.765 \\
\hline
SG  & M0 & 2 & 417.709 & 839.417 & 844.706 \\
    & M1 & 3 & \textbf{401.342} & \textbf{808.684} & \textbf{816.618} \\
    & M3 & 5 & 416.177 & 842.354 & 855.576 \\
\hline
IG  & M0 & 2 & 182.382 & 368.763 & \textbf{374.052} \\
    & M1 & 3 & 180.781 & \textbf{367.563} & 375.496 \\
    & M3 & 5 & \textbf{180.113} & 370.226 & 383.448 \\
\hline
\end{tabular}
\end{table}

M3 improves the maximized likelihood relative to M0 in all three classes, as
expected from the added flexibility. The reduction in NLL is nevertheless small:
approximately $1.09$ for ALL, $1.53$ for SG, and $2.27$ for IG. After accounting
for the three additional parameters, M3 does not outperform M0 under either AIC
or BIC. More importantly, for ALL and SG the improvement from M1 is much larger:
relative to M0, M1 lowers the NLL by approximately $17.54$ and $16.37$,
respectively. Thus, for these two classes, the Davis--Lo likelihood gain is
robust to replacing the single-Gaussian environmental distribution with this
more flexible two-component alternative.

The IG class gives a weaker result. M3 attains the lowest NLL, but its additional
parameters are not supported by AIC or BIC, while M1 is favored only slightly by
AIC and M0 by BIC. The annual IG data therefore do not provide the same strength
of evidence for an additional contagion component as the ALL and SG data.
Accordingly, this robustness test does not imply identification against arbitrary
forms of environmental heterogeneity; it shows that the strong M1 improvement
for ALL and SG is not explained by the two-component Gaussian-mixture null
considered here.

The corresponding best-found M3 parameter estimates are reported in
Appendix~\ref{app:m3_parameters}.
}


\section{Discussion}
\label{sec:conclusion}
This paper examined the identifiability of contagion in aggregated annual
default counts by comparing three representative frameworks: the Vasicek
common-factor model, the Davis--Lo cumulative-contagion model, and the Torri
threshold-type contagion model.

Under {the static benchmark}, the Vasicek model provides the best overall
fit, particularly in reproducing the observed tail behavior.
{
Because the Vasicek default-count likelihood is exactly equivalent to a
Probit--Normal binomial mixture, this result does not by itself identify a
common-factor mechanism. Rather, it shows that a smooth environmental mixture
can absorb a substantial part of the variation in aggregated annual default
counts.
}

To separate cross-year heterogeneity from within-year contagion, we
{
used the Probit--Normal mixture as an environmental null model (M0) and
introduced the Davis--Lo and Torri contagion mechanisms as extensions M1 and
M2, respectively. 
The variance decomposition reported in Appendix~\ref{app:hier_additional} shows that fluctuations in the annual default probability provide the largest normalized contribution to aggregate variability, while the contribution of contagion is substantially smaller and strongly mechanism- and class-dependent.
For ALL and SG, M1 provides a substantial improvement over M0 that survives
both AIC and BIC penalties. In contrast, M2 collapses to the environmental
null for ALL and IG and, although non-trivial for SG, remains clearly inferior
to M1. Evidence for an additional contagion component is much weaker for IG.
}

The comparison of tail behavior further highlights this distinction.
While the Vasicek model performs best under the {static benchmark},
{
M1 provides the closest overall description of the upper tail for ALL and SG
once environmental fluctuations are introduced.
For ALL and IG, M2 coincides with M0, whereas for SG it remains distinct
and produces a heavier upper tail than M0.
}
{
Thus, the likelihood improvement associated with cumulative contagion in ALL
and SG is also visible in the shape of the default-count distribution and is
not confined to small adjustments near its center.
}

{
Taken together with the variance decomposition in Appendix C, these results show that, at the annual observation scale, most of the variation in aggregated default counts is represented by cross-year fluctuations in the latent default probability, whereas the residual within-year contagion component identified conditional on this variation is substantially smaller. This scale-dependent decomposition should not be interpreted as evidence that microscopic interactions are unimportant: the dynamics of the latent default probability itself may reflect mechanisms operating at finer temporal scales, including contagion, self-excitation, network interactions, or common environmental forces.}
Contagion, if present, appears to refine the tail shape rather than serve as
the main source of variance { in the annual aggregated counts.
From the perspective of identifiability, threshold-type contagion is more
readily absorbed by environmental fluctuations and does not generate a distinct
component for ALL and IG, whereas cumulative contagion leaves a detectable
distributional contribution for ALL and SG that remains distinguishable after
aggregation.}
This distinction suggests that identifiability depends on whether a mechanism leaves a detectable structural signature after aggregation, rather than simply on whether it generates dependence at the microscopic level.

{
The robustness analysis strengthens this interpretation within the class of
environmental nulls considered here. Replacing the single Gaussian latent
environment by a two-component Gaussian mixture (M3) produces only modest
likelihood improvements over M0 and is not favored after accounting for model
complexity. For ALL and SG, the improvement from M1 remains substantially
larger. This result does not imply identification against arbitrarily flexible
forms of environmental heterogeneity; it shows that the M1 advantage is not
explained by the two-component Gaussian-mixture alternative considered here.}

The {environmental-null} specification used here is deliberately
conservative: by allowing the baseline default probability to fluctuate
independently across years, it gives environmental heterogeneity substantial
freedom to absorb variation that might otherwise be attributed to contagion.
The remaining ambiguity is not intrinsic to the models alone, but also
reflects the limited temporal resolution of annual data.
When higher-frequency observations are available, temporal information can
provide an additional source of identification by modeling the latent
default-probability path dynamically and using its correlation structure to
separate persistent environmental fluctuations from residual contagion or
same-period correlation {\cite{Mori2026TemporalCoarseGraining}}.
Such a dynamical coarse-graining approach provides a complementary extension
of the static {distributional} analysis developed here.

{ More realistic network-based contagion models constitute another natural
extension. Empirical studies of Japanese inter-firm trading and shareholding
networks have revealed heterogeneous network structures and have also examined
the propagation of bankruptcy risk through actual inter-firm connections.
Such empirical network structures may provide useful guidance for constructing
more realistic bankruptcy-contagion models in future applications
\cite{GotoTakayasuTakayasu2017,SoumaFujiwaraAoyama2005}.}

{
Overall, the results indicate that identifiability from coarse-grained default
counts depends not only on the form of the underlying mechanism and the
flexibility of the environmental null, but also on the amount of information
available in the data, including the sample size, event frequency, and
observation scale.
}

\section*{Acknowledgments}

This research was supported by JSPS KAKENHI Grant Number JP26K06955.

During the preparation of this work, the author used ChatGPT, an AI-assisted language model developed by OpenAI, to assist with manuscript drafting, rewriting, language editing, structural revision, and the generation and refinement of data-analysis and visualization code. After using this tool, the author reviewed and edited the manuscript, checked the analysis, numerical results, and figures, and takes full responsibility for the content of the manuscript.

\newpage

\appendix

\section{One-period model properties and distributional relations}
\label{app:models}

In this appendix, we derive the mean default probability and pairwise dependence measures for the Davis--Lo and Torri infectious default models. These expressions clarify the structural distinction between cumulative and threshold-type contagion and support the supplementary comparisons reported below. The same homogeneous-portfolio assumption as in the main text is used throughout.

\subsection{Davis--Lo model}
\label{app:davis-lo}

In the Davis--Lo model, the final default indicator is given by
\begin{equation}
Z_i
=
X_i + (1-X_i)
\left(
1-\prod_{j\neq i}(1-Y_{ij}X_j)
\right),
\end{equation}
where
\begin{equation}
X_i \sim \mathrm{Bernoulli}(p), \qquad
Y_{ij} \sim \mathrm{Bernoulli}(q),
\end{equation}
and all variables are independent.

We define the contagion indicator
\begin{equation}
I_{i,n}^C
:=
1-\prod_{j\neq i}(1-Y_{ij}X_j),
\end{equation}
so that
\[
Z_i = X_i + (1-X_i) I_{i,n}^C.
\]

Let $K = \sum_{i=1}^n X_i$ denote the number of idiosyncratic defaults. Conditional on $K=k$, each surviving obligor defaults independently with probability
\[
r_k = 1 - (1-q)^k.
\]
Hence,
\[
L_n \mid K=k \sim k + \mathrm{Binomial}(n-k, r_k).
\]

\subsubsection{Moment generating function}

The moment generating function (MGF) of $L_n$ can be written as a mixture over $K$:
\begin{equation}
M_{L_n}(t)
=
\sum_{k=0}^n
P(K=k)\,
e^{tk}
\left[(1-r_k)+r_k e^t\right]^{n-k}.
\end{equation}
This expression shows that the Davis--Lo default-count distribution is a finite mixture
of shifted binomial distributions indexed by the number of idiosyncratic seed defaults $K$.
The dependence among final defaults is generated through the $k$-dependent contagion
probability $r_k=1-(1-q)^k$, which increases with the number of seed defaults.

\subsubsection{Mean}

For a given obligor $i$, define
\[
K_{-i}=\sum_{j\neq i}X_j .
\]
Conditional on $X_i=0$, infection of obligor $i$ is generated by the $K_{-i}$ idiosyncratic defaults among the other $n-1$ obligors. Therefore,
\[
\mathbb P(Z_i=1)
=
p+(1-p)\mathbb E\left[1-(1-q)^{K_{-i}}\right].
\]
Since $K_{-i}\sim \mathrm{Binomial}(n-1,p)$,
\[
\mathbb E[(1-q)^{K_{-i}}]=(1-pq)^{n-1}.
\]
Thus, the mean default probability is
\[
m
=
p+(1-p)\left[1-(1-pq)^{n-1}\right].
\]

\subsubsection{Covariance}

A closed-form expression for the covariance is obtained as
\[
\operatorname{Cov}(Z_i,Z_j)
=
(1-p)^2
\Bigl[
(1-2pq+pq^2)^{n-2}
-
(1-pq)^{2(n-1)}
\Bigr].
\]
The pairwise correlation is
\[
\rho=\frac{\operatorname{Cov}(Z_i,Z_j)}{m(1-m)}.
\]

\subsection{Torri model (infection with immunization)}
\label{app:torri}

In the Torri model, the final default indicator is
\begin{equation}
Z_i=X_i + (1-X_i)(1-U_i)
\left(
1-\prod_{j=1}^n (1-X_jV_j)
\right),
\end{equation}
where
\[
X_i \sim \mathrm{Bernoulli}(p),\quad
U_i \sim \mathrm{Bernoulli}(u),\quad
V_i \sim \mathrm{Bernoulli}(v),
\]
and all variables are independent. The product may equivalently be restricted to $j\ne i$, since the contagion term is multiplied by $1-X_i$.

Define the global contagion indicator
\begin{equation}
I_n^C:=
\mathbf{1}\left(\sum_{j=1}^n X_jV_j > 0\right).
\end{equation}
Then
\[
Z_i = X_i + (1-X_i)(1-U_i)I_n^C.
\]

It is instructive to relate this model to the Davis--Lo framework. In the Davis--Lo model, contagion is governed by pairwise indicators $Y_{ij}$. If these are factorized as
\[
Y_{ij} = [1 - U_i] V_j,
\]
then substituting into the Davis--Lo contagion term yields exactly the above formulation. 
In this sense, the Torri model can be interpreted as a constrained version
of the Davis--Lo model, in which cumulative contagion is replaced by a
global threshold-type mechanism activated by the existence of at least one
infectious default.

\subsubsection{Moment generating function}

Let $L_n = \sum_{i=1}^n Z_i$ denote the total number of defaults.
The moment generating function (MGF) admits a simple representation:
\begin{equation}
M_{L_n}(t)=\left(a e^t + b + c\right)^n+\left((p + b)e^t + c\right)^n-
\left((a + b)e^t + c\right)^n,
\end{equation}
where
\[
a = p(1 - v), \quad
b = (1 - p)(1 - u), \quad
c = (1 - p)u.
\]
This representation shows that the Torri model also has a binomial-mixture-type structure, but with a fundamentally different dependence mechanism from that of the Davis--Lo model.

\subsubsection{Mean}

From
\[
Z_i = X_i + (1-X_i)(1-U_i) I_n^C,
\]
we have
\begin{align*}
\mathbb E[Z_i]
&= \mathbb E[X_i] + \mathbb E[(1-X_i)(1-U_i) I_n^C] \\
&= p + \mathbb E[(1-X_i)(1-U_i) I_n^C].
\end{align*}
Now, when $X_i=0$, the contagion indicator $I_n^C$ is triggered only by the remaining $n-1$ obligors, so
\[
\Pr(I_n^C=1 \mid X_i=0) = \pi_{n-1},
\]
where $\pi_{n-1}=1-(1-pv)^{n-1}$
is the probability that, 
for a given obligor $i$, the contagious state is activated by at least one of the other $n-1$ obligors.

Using the independence of $U_i$ from $(X_j,V_j)_{j \neq i}$, we obtain
\begin{align*}
\mathbb E[(1-X_i)(1-U_i) I_n^C]
&= \Pr(X_i=0)\Pr(U_i=0)\Pr(I_n^C=1 \mid X_i=0) \\
&= (1-p)(1-u)\pi_{n-1}.
\end{align*}
Therefore,
\[
m = p + (1-p)(1-u)\pi_{n-1}.
\]

\subsubsection{Pairwise correlation}

We now derive the joint default probability $E[Z_i Z_j]$ for two distinct obligors $i \neq j$.

Recall that
\begin{align*}
Z_i &= X_i + (1-X_i)(1-U_i) I_n^C, \\
Z_j &= X_j + (1-X_j)(1-U_j) I_n^C,
\end{align*}
where
\[
I_n^C = \mathbf{1}\!\left( \sum_{k=1}^n X_k V_k > 0 \right).
\]

Thus, for $i \neq j$,
\begin{align*}    
Z_i Z_j
&=
\Bigl[X_i + (1-X_i)(1-U_i) I_n^C\Bigr]
\Bigl[X_j + (1-X_j)(1-U_j) I_n^C\Bigr] \\
&=
X_i X_j
+ X_i (1-X_j)(1-U_j) I_n^C
+ X_j (1-X_i)(1-U_i) I_n^C \nonumber\\
&\quad
+ (1-X_i)(1-U_i)(1-X_j)(1-U_j) (I_n^C)^2.
\end{align*}
Since $(I_n^C)^2 = I_n^C$, this becomes
\begin{align*}
Z_i Z_j
&=
X_i X_j
+ X_i (1-X_j)(1-U_j) I_n^C
+ X_j (1-X_i)(1-U_i) I_n^C \nonumber\\
&\quad
+ (1-X_i)(1-U_i)(1-X_j)(1-U_j) I_n^C.
\end{align*}
Taking expectations, we obtain
\begin{align*}
E[Z_i Z_j]
&= E[X_i X_j]
+ E\!\left[X_i (1-X_j)(1-U_j) I_n^C\right]
+ E\!\left[X_j (1-X_i)(1-U_i) I_n^C\right] \nonumber\\
&\quad
+ E\!\left[(1-X_i)(1-U_i)(1-X_j)(1-U_j) I_n^C\right].
\end{align*}
By symmetry, the second and third terms are equal.

First,
\[
E[X_i X_j] = P(X_i=1, X_j=1) = p^2.
\]

Next, consider
\[
E\!\left[X_i (1-X_j)(1-U_j) I_n^C\right].
\]
This term is nonzero only when $X_i=1$, $X_j=0$, and $U_j=0$, so
\[
E\!\left[X_i (1-X_j)(1-U_j) I_n^C\right]
=
p(1-p)(1-u)\,
P(I_n^C=1 \mid X_i=1, X_j=0).
\]
Under $X_i=1$ and $X_j=0$, contagion fails to occur only if $V_i=0$ and no other obligor $k \neq i,j$ satisfies $X_k V_k = 1$. Hence
\begin{align*}
P(I_n^C=0 \mid X_i=1, X_j=0)
&=
P(V_i=0)\prod_{k \neq i,j} P(X_k V_k = 0) \\
&=
(1-v)(1-pv)^{n-2}.
\end{align*}
Therefore,
\[
P(I_n^C=1 \mid X_i=1, X_j=0)=1 - (1-v)(1-pv)^{n-2},
\]
and thus
\[
E\!\left[X_i (1-X_j)(1-U_j) I_n^C\right]
=
p(1-p)(1-u)\Bigl[1-(1-v)(1-pv)^{n-2}\Bigr].
\]
By symmetry,
\[
E\!\left[X_j (1-X_i)(1-U_i) I_n^C\right]
=
p(1-p)(1-u)\Bigl[1-(1-v)(1-pv)^{n-2}\Bigr].
\]

Finally, consider
\[
E\!\left[(1-X_i)(1-U_i)(1-X_j)(1-U_j) I_n^C\right].
\]
This is nonzero only when $X_i=0$, $X_j=0$, $U_i=0$, and $U_j=0$, so
\[
E\!\left[(1-X_i)(1-U_i)(1-X_j)(1-U_j) I_n^C\right]
=
(1-p)^2(1-u)^2\,
P(I_n^C=1 \mid X_i=0, X_j=0).
\]
Under $X_i=0$ and $X_j=0$, contagion can be triggered only by one of the remaining $n-2$ obligors. Thus
\begin{align*}
P(I_n^C=0 \mid X_i=0, X_j=0)
&=
\prod_{k \neq i,j} P(X_k V_k = 0) \\
&=
(1-pv)^{n-2},
\end{align*}
so that
\[
P(I_n^C=1 \mid X_i=0, X_j=0)=1 - (1-pv)^{n-2}.
\]
Therefore,
\[
E\!\left[(1-X_i)(1-U_i)(1-X_j)(1-U_j) I_n^C\right]
=(1-p)^2(1-u)^2 \Bigl[1-(1-pv)^{n-2}\Bigr].
\]

Combining all terms, we obtain
\begin{align*}
E[Z_i Z_j]
&=
p^2
+ 2p(1-p)(1-u)\Bigl[1-(1-v)(1-pv)^{n-2}\Bigr] \nonumber\\
&\quad
+ (1-p)^2(1-u)^2\Bigl[1-(1-pv)^{n-2}\Bigr].
\end{align*}

Hence,
\[
\operatorname{Cov}(Z_i,Z_j)
=
E[Z_i Z_j] - m^2,
\]
where $m = E[Z_i]$.
Finally, the pairwise correlation is given by
\[
\rho=\frac{\operatorname{Cov}(Z_i,Z_j)}{m(1-m)}.
\]
This derivation shows that dependence in the Torri model is generated by the common binary contagion state $I_n^C$. Unlike the Davis--Lo model, the contagion effect does not accumulate with the number of infectious obligors; only the existence of at least one infectious trigger matters.

This structural difference underlies the identifiability results in the main text.

\subsection{Activation probability and tail-risk trade-off in the Torri model}
\label{app:torri-2}

To illustrate the residual degree of freedom in the Torri model, we examine how variation in the parameter $p$ along the iso-$(m,\rho)$ manifold affects both the activation probability $\pi_n$ and the resulting tail risk. 
Although the activation probability
\[
\pi_n = 1 - (1 - p v)^n
\]
is increasing in $p$ for fixed $v$, the constraint of constant $(m,\rho)$ induces a compensating decrease in $v$, leading to a decreasing relationship between $p$ and $\pi_n$ along the iso-curve.

\begin{figure}[t]
\centering

\begin{minipage}{0.42\textwidth}
\centering
\includegraphics[width=\linewidth]{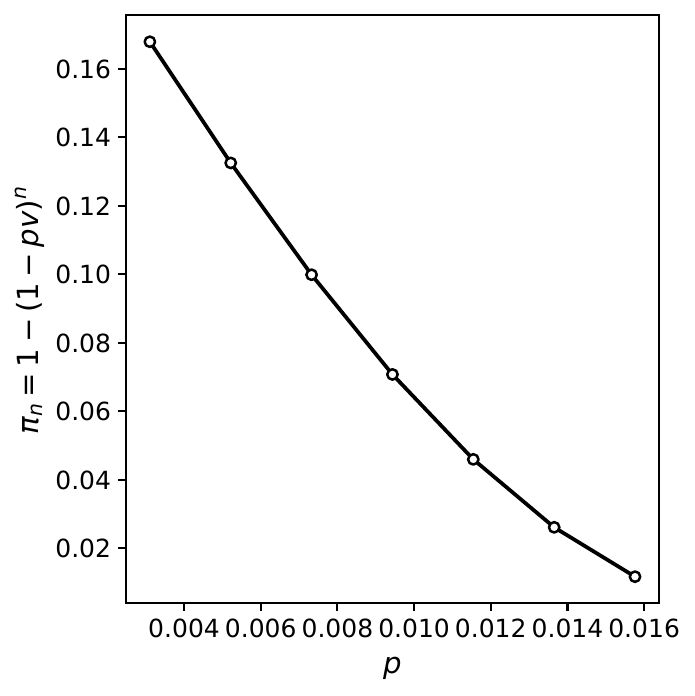}
\end{minipage}
\hfill
\begin{minipage}{0.54\textwidth}
\centering
\includegraphics[width=\linewidth]{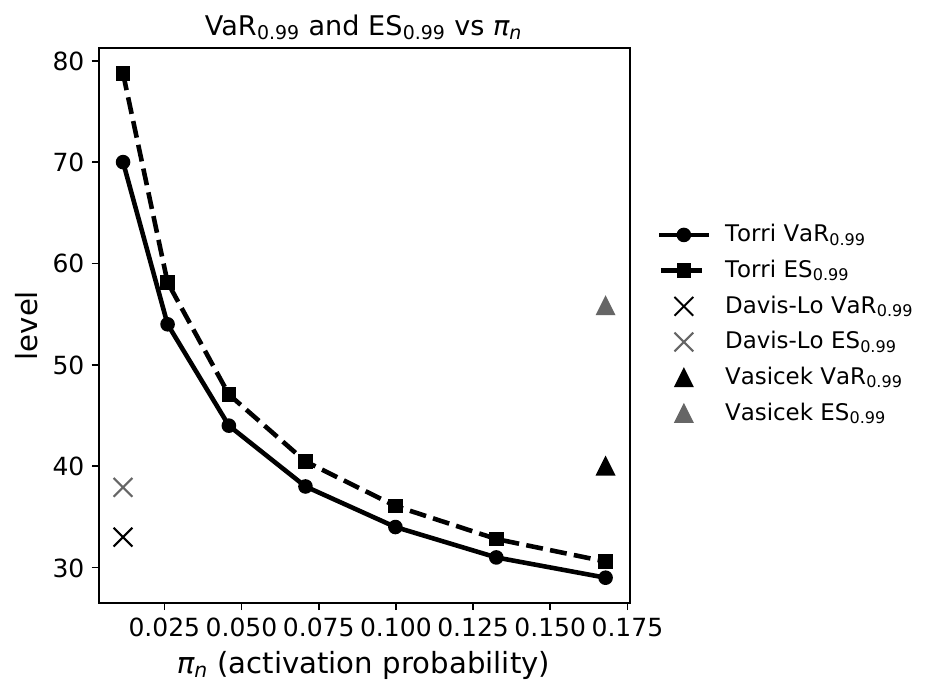}
\end{minipage}
\caption{
Illustration of the residual degree of freedom in the Torri model under fixed $(m,\rho)$.
\textbf{Left:} Activation probability $\pi_n = 1 - (1 - pv)^n$ as a function of $p$ along the iso-$(m,\rho)$ manifold.
\textbf{Right:} Dependence of $\mathrm{VaR}_{0.99}$ and $\mathrm{ES}_{0.99}$ on $\pi_n$ for $n=200$, $m=0.02$, and $\rho=0.08$.
Together, the panels show how variation in $p$ translates into changes in $\pi_n$, which in turn determines the frequency--severity trade-off in contagion events.
}
\label{fig:torri_mechanism}
\end{figure}

Figure~\ref{fig:torri_mechanism} summarizes this mechanism. 
The left panel shows how $\pi_n$ varies with $p$ along the iso-$(m,\rho)$ manifold. 
The right panel shows how the corresponding values of $\pi_n$ translate into tail-risk measures, specifically $\mathrm{VaR}_{0.99}$ and $\mathrm{ES}_{0.99}$.

Within the Torri family, both $\mathrm{VaR}_{0.99}$ and $\mathrm{ES}_{0.99}$ decrease over the range shown as $\pi_n$ increases. 
Hence, smaller values of $\pi_n$ correspond to rarer but more severe contagion events, whereas larger values of $\pi_n$ correspond to more frequent but milder cascades.

The Davis--Lo and Vasicek models appear as single points in this diagram, since their dependence structures are fully determined once $(m,\rho)$ is fixed. 
By contrast, the Torri model generates a continuous range of tail risks even under identical first and second moments. 
Together, the two panels provide a compact representation of the frequency--severity trade-off underlying the Torri mechanism.

\subsection{Vasicek model}
\label{app:vasicek}

The moment generating function (MGF) can be written as
\begin{equation}
M_{L_n}(t) = \mathbb{E}\left[\left((1-p(F)) + p(F)e^{t}\right)^{n}\right],
\end{equation}
where the expectation is taken with respect to $F$. 

The mean default probability is
\begin{equation}
m = p.
\end{equation}
The joint default probability is given by
\begin{equation}
\mathbb{P}(Z_i=1, Z_j=1)
=\Phi_2\!\left(\Phi^{-1}(p),\,\Phi^{-1}(p);\,\rho_A\right).
\end{equation}
Here, $\Phi_2(\cdot,\cdot;\rho_A)$ denotes the cumulative distribution function of 
the bivariate standard normal distribution with correlation parameter $\rho_A$.
Hence the pairwise \emph{default correlation} is
\begin{equation}
\rho
=
\frac{
\Phi_2\!\left(\Phi^{-1}(p),\,\Phi^{-1}(p);\,\rho_A\right) - p^2
}{
p(1-p)
}.
\end{equation}
Thus, in the Vasicek model, dependence is induced by the latent asset correlation parameter $\rho_A$, while the observable default correlation $\rho$ is a nonlinear function of both $p$ and $\rho_A$.

\subsection{Distributional relations among the static one-period model families}
\label{subsec:kl-iid}

The Davis--Lo, Torri, and Vasicek models generate dependence through
different one-period mechanisms. Even when the mean default probability
$m$ and pairwise default correlation $\rho$ are matched, these models
need not generate the same full default-count distribution. 
We summarize
this structural separation 
using the { directed
Kullback--Leibler (KL) divergence \cite{KullbackLeibler1951,CoverThomas2006}.}

For two probability mass functions $P$ and $Q$ on
$\{0,1,\ldots,n\}$, the KL divergence is defined by
\[
D_{\mathrm{KL}}(P\|Q)
=
\sum_{h=0}^{n}
P(h)\log\frac{P(h)}{Q(h)}.
\]
Because the KL divergence is asymmetric, the approximation of one model
family by another depends on which distribution is taken as the target.
For a target distribution $P_{\mathrm{target}}$ and a competing model
family $\mathcal{M}$, we define the KL projection
\[
D_{\mathrm{KL}}(P_{\mathrm{target}}\|\mathcal{M})
=
\inf_{\theta\in\Theta_{\mathcal{M}}}
D_{\mathrm{KL}}
\left(
P_{\mathrm{target}}
\|P_{\mathcal{M}}(\cdot;\theta)
\right).
\]
A small projection divergence indicates that the competing family can
closely approximate the target count distribution, whereas a positive
gap indicates residual distributional structure not reproduced by that
family.

Table~\ref{tab:kl_iid} reports the minimum KL projection divergences for
the representative calibration
$n=200$, $m=0.02$, and $\rho=0.08$.
The target distributions are the Davis--Lo model and three Torri
specifications corresponding to high, intermediate, and low activation
probabilities $\pi_n$. The three Torri targets differ along the
iso-$(m,\rho)$ manifold discussed in the preceding subsection.

\begin{table}[t]
\centering
\caption{
Minimum KL projection divergences from representative static one-period 
default-count distributions to alternative model families under
$n=200$, $m=0.02$, and $\rho=0.08$.
Each row denotes the target distribution and each column gives the
minimum KL divergence to the indicated model family.
}
\begin{tabular}{lccc}
\hline
Target distribution
& Best-fit Torri
& Best-fit Davis--Lo
& Best-fit Vasicek \\
\hline
Torri (high $\pi_n$)
& 0.000 & 1.233 & 0.305 \\
Torri (mid $\pi_n$)
& 0.000 & 1.172 & 0.368 \\
Torri (low $\pi_n$)
& 0.000 & 0.273 & 0.273 \\
Davis--Lo
& 0.096 & 0.000 & 0.258 \\
\hline
\end{tabular}
\label{tab:kl_iid}
\end{table}

Matching $m$ and $\rho$ therefore does not, in general, eliminate the
distributional differences among the one-period mechanisms. The
separation is nevertheless asymmetric and parameter dependent. The
high- and intermediate-$\pi_n$ Torri distributions are well separated
from the Davis--Lo family, whereas the low-$\pi_n$ branch lies closer to
both the Davis--Lo and Vasicek families. Conversely, the Davis--Lo
target is approximated more closely by the Torri family than by the
Vasicek family, although the minimum divergence remains positive.

The Vasicek family provides a comparatively close approximation to the
infectious target distributions despite its different construction.
This illustrates an important feature of aggregated default counts:
distinct one-period dependence mechanisms can generate similar count
distributions after coarse graining. At the same time, the nonzero KL
projection divergences show that matching the first two dependence
summaries does not generally remove all information contained in the
full distribution.

The KL projections in this subsection are used only to characterize the
distributional relations among the static one-period model families. They are not
used as an additional model-selection criterion and do not imply
identification of a unique microscopic mechanism.

\newpage

\section{Additional Empirical Results for the Static Setting}
\label{app:add_iid}

\subsection{Computation of aggregated moments}

Because the pool size $n_t$ varies across years, the model-implied mean default probability $m$ and joint default probability
$\mathbb{P}(Z_i=1, Z_j=1)$
are first evaluated at each observed $n_t$ and then aggregated using the same weighting scheme as in the data:
\[
m = \frac{\sum_t n_t\, m(n_t)}{\sum_t n_t},
\qquad
\mathbb{P}(Z_i=1, Z_j=1) =
\frac{\sum_t n_t(n_t-1)\,\mathbb{P}(Z_i=1, Z_j=1\mid n_t)}
{\sum_t n_t(n_t-1)}.
\]
The implied pairwise correlation is then defined by
\[
\rho = \frac{\mathbb{P}(Z_i=1, Z_j=1)-m^2}{m(1-m)}.
\]

{
The empirical counterparts are computed directly from the annual counts. Specifically,
\[
\hat m
=
\frac{\sum_t L_t}{\sum_t n_t},
\qquad
\widehat{P}_{2}
=
\frac{\sum_t L_t(L_t-1)}
     {\sum_t n_t(n_t-1)},
\]
and the empirical pairwise default correlation is
\[
\hat\rho
=
\frac{\widehat{P}_{2}-\hat m^2}
     {\hat m(1-\hat m)}.
\]
Thus, the empirical pairwise correlation is inferred from the first two factorial moments of the aggregated annual default counts under the homogeneous exchangeability assumption; individual default-pair identities are not required.}

\subsection{Additional distributional results for the SG class}

\begin{figure}[htbp]
\centering
\begin{minipage}{0.48\textwidth}
\centering
\includegraphics[width=\linewidth]{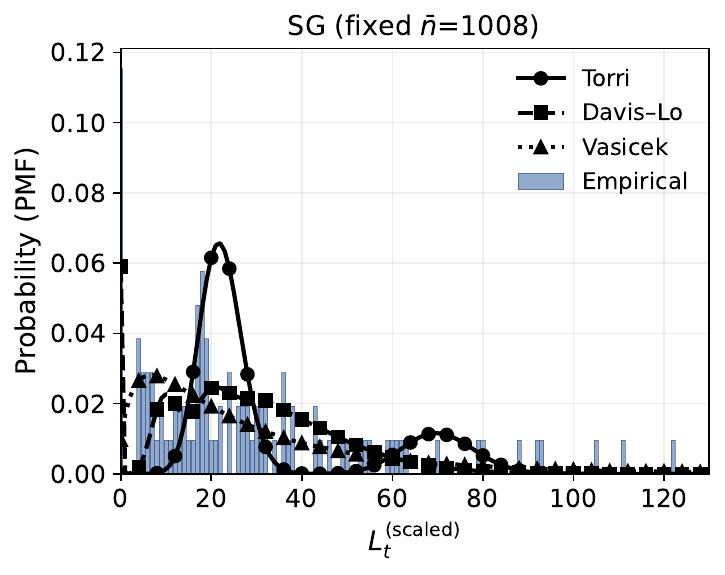}
\end{minipage}
\hfill
\begin{minipage}{0.48\textwidth}
\centering
\includegraphics[width=\linewidth]{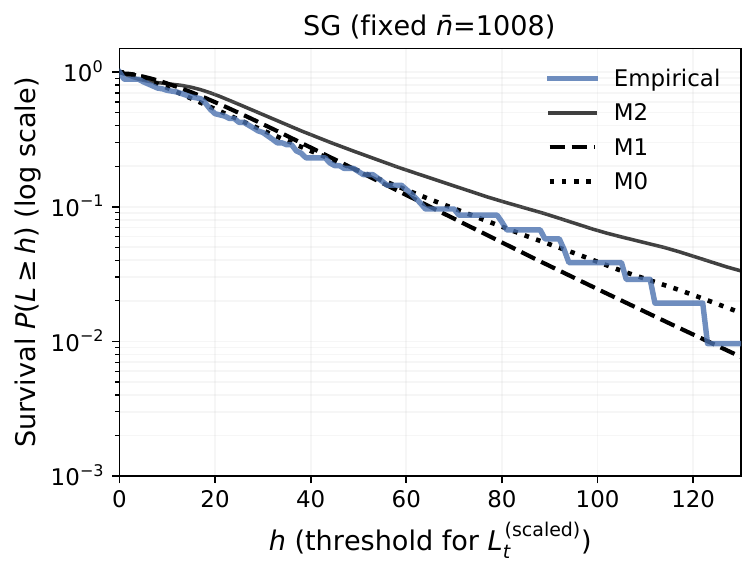}
\end{minipage}

\caption{
Comparison of model-implied distributions and tail behavior at the fixed portfolio size $n=\bar n$ for the SG class.
\textbf{Left:} Probability mass functions (PMFs).
\textbf{Right:} Survival functions $S(h)=P(L \ge h)$ on a logarithmic scale.
The Torri model exhibits a segmented PMF shape and rapidly decaying tails, the Davis--Lo model provides a smoother but still insufficient decay, and the Vasicek model provides a closer match to the empirical tail behavior over a broad range of thresholds.
}
\label{fig:pmf_survival_sg}
\end{figure}

In the main text, we report the PMF and survival function for the ALL and IG portfolios.
For completeness, we report the corresponding results for the SG portfolio in this appendix.
Figure~\ref{fig:pmf_survival_sg} reports the corresponding PMFs and survival functions for the SG class at the fixed portfolio size $n=\bar n$.
The same qualitative pattern emerges as in the ALL and IG classes: the Torri model generates a segmented distribution with thin tails, the Davis--Lo model improves smoothness but still understates tail risk, whereas the Vasicek model remains closest to the empirical distribution, especially in the tail region.

\newpage

\section{Additional Results for the Environmental-Null Extensions}
\label{app:hier_additional}

\begin{figure}[htbp]
\centering

\begin{minipage}{0.48\textwidth}
\centering
\includegraphics[width=\linewidth]{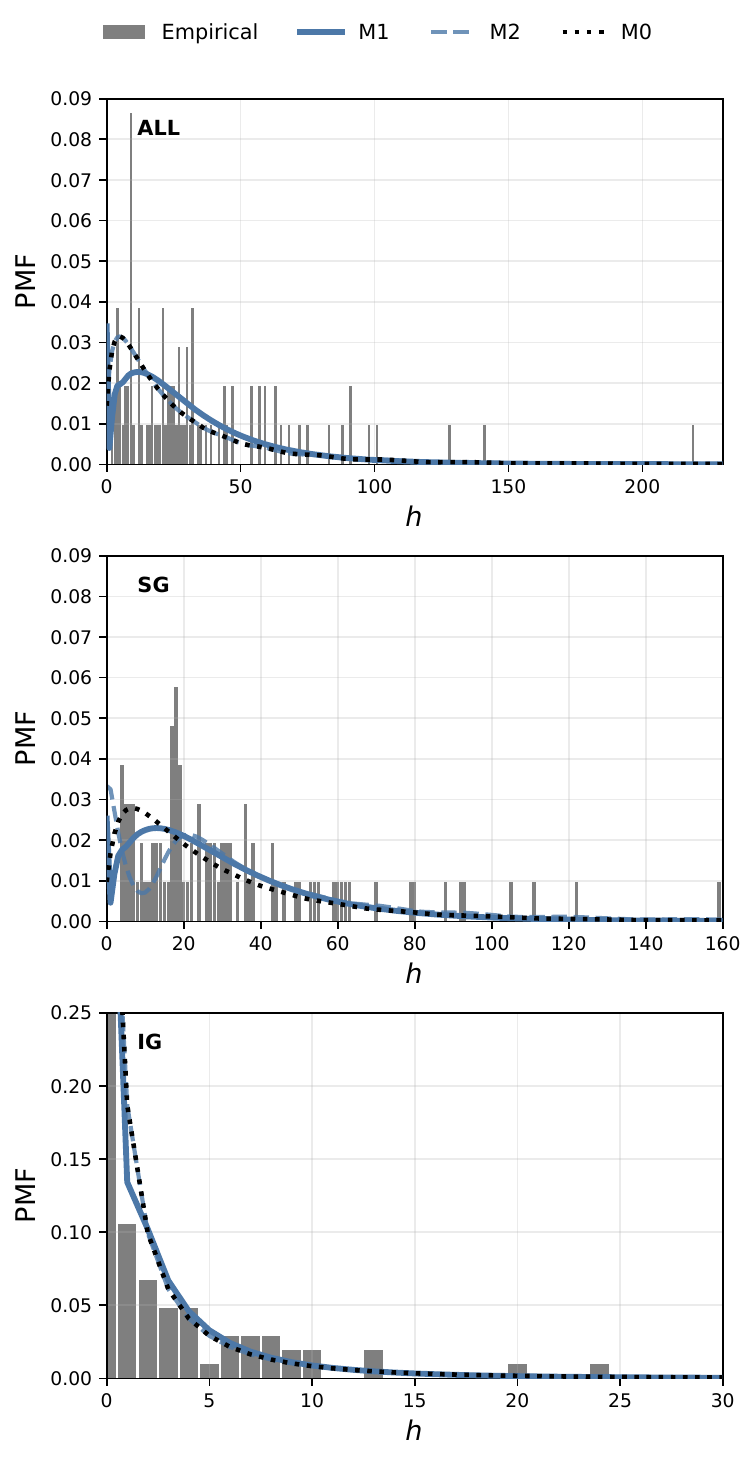}
\end{minipage}
\hfill
\begin{minipage}{0.48\textwidth}
\centering
\includegraphics[width=\linewidth]{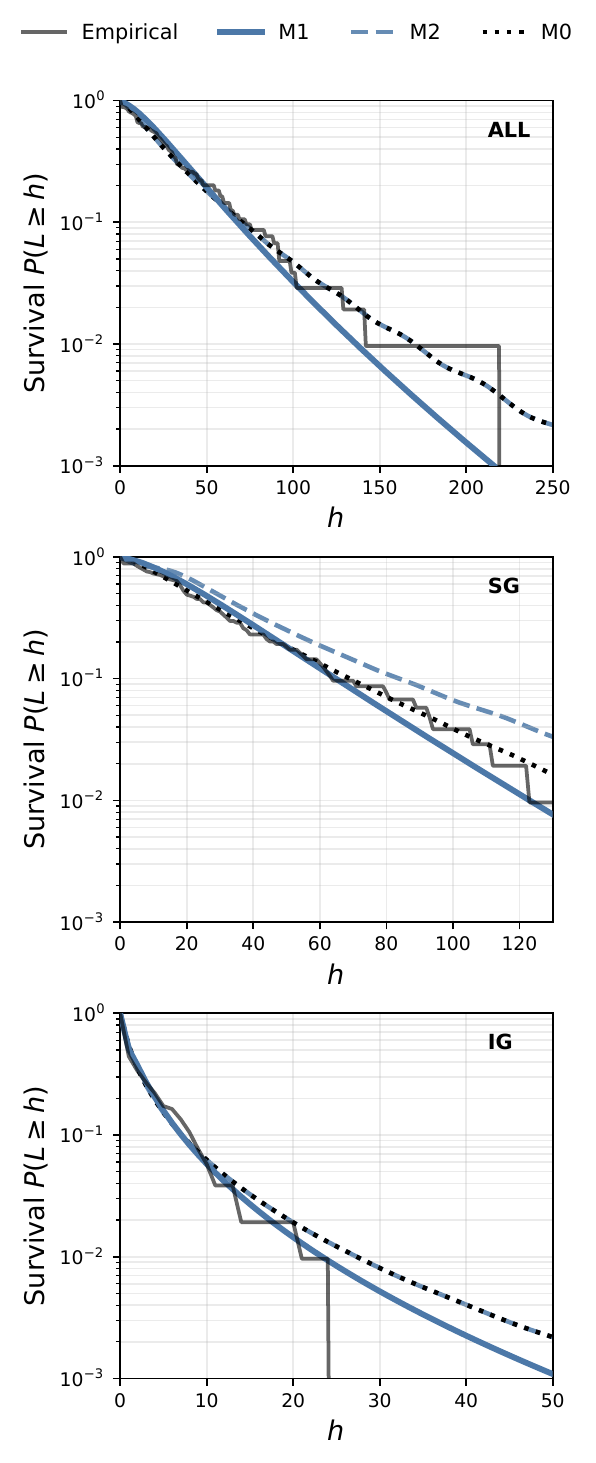}
\end{minipage}

\caption{
Additional fits of the environmental null and its contagion extensions across
classes.
\textbf{Left:} Probability mass functions.
\textbf{Right:} Survival functions $P(L \ge h)$ on a logarithmic scale.
M1 is closest to the empirical distributions.
{ For ALL and IG, M2 coincides with M0, whereas for SG M2 remains distinct
and produces a heavier upper tail than M0.}
}
\label{fig:pmf_survival_m012_appendix}
\end{figure}

For completeness, Figure~\ref{fig:pmf_survival_m012_appendix} reports the PMFs
and class-specific survival functions for the environmental null and its
contagion extensions across all three credit classes.
Overall, M1 provides the closest fit to the empirical distributions.
{ For ALL and IG, M2 coincides with M0, whereas for SG M2 remains distinct
and produces a heavier upper tail than M0.}

\subsection{Full-sample variance decomposition}
\label{app:variance_full}

As a supplementary diagnostic, we decompose the model-implied variance of the
scaled annual default count into three components under M1 and M2: the
conditional binomial contribution, the additional within-period
contagion contribution, and the cross-year contribution induced by
fluctuations in the baseline default probability $p_t$. The decomposition is
evaluated at the fixed portfolio size $\bar n$ used for the distributional
comparisons.

For each contagion extension, let
\begin{equation}
m_t
=
\frac{E[L_t\mid p_t,\theta]}{\bar n}
\end{equation}
denote the conditional final default probability. We define
\begin{align}
R_{\mathrm{iid}}
&=
\frac{
E_{p_t}[\bar n m_t(1-m_t)]
}{
\operatorname{Var}_{\mathrm{data}}(L_t^{\mathrm{scaled}})
},
\\
R_{\mathrm{infect}}
&=
\frac{
E_{p_t}
[
\operatorname{Var}(L_t\mid p_t,\theta)
-\bar n m_t(1-m_t)
]
}{
\operatorname{Var}_{\mathrm{data}}(L_t^{\mathrm{scaled}})
},
\\
R_{p_t}
&=
\frac{
\operatorname{Var}_{p_t}(E[L_t\mid p_t,\theta])
}{
\operatorname{Var}_{\mathrm{data}}(L_t^{\mathrm{scaled}})
}.
\end{align}

\begin{table}[htbp]
\centering
\caption{
Variance decomposition relative to the empirical variance of the scaled annual
default counts. The entries are normalized contributions, not shares of the
model variance.
}
\label{tab:variance_decomp_full}
\begin{tabular}{lcccccc}
\hline
& \multicolumn{3}{c}{M1 (Davis--Lo)}
& \multicolumn{3}{c}{M2 (Torri)}\\
Class
& $R_{\mathrm{iid}}$
& $R_{\mathrm{infect}}$
& $R_{p_t}$
& $R_{\mathrm{iid}}$
& $R_{\mathrm{infect}}$
& $R_{p_t}$\\
\hline
ALL & 0.026 & 0.090 & 0.575 & 0.025 & 0.000 & 0.998\\
SG  & 0.034 & 0.103 & 0.648 & 0.041 & 0.019 & 1.666\\
IG  & 0.137 & 0.101 & 1.175 & 0.145 & 0.000 & 1.826\\
\hline
\end{tabular}
\end{table}

{
Because the components are normalized by the empirical variance rather than
by the total model-implied variance, they need not sum to one.
The largest normalized contribution in both specifications is associated with
fluctuations in $p_t$, indicating that, at the annual observation scale,
cross-year variation in the latent default probability represents a large
component of the observed variation in annual default counts.} At the same time, the two contagion mechanisms differ.
M1 retains a {positive} infection contribution in all three
classes, whereas the residual infection contribution under M2 is essentially
zero for ALL and IG and remains small for SG. This supplementary decomposition
is consistent with the likelihood-based identification results in
Sec.~\ref{sec:identification}: threshold-type contagion is more readily
absorbed by environmental heterogeneity, while cumulative contagion leaves a
residual component after aggregation.

\begin{figure}[htbp]
\centering
\includegraphics[width=0.8\linewidth]{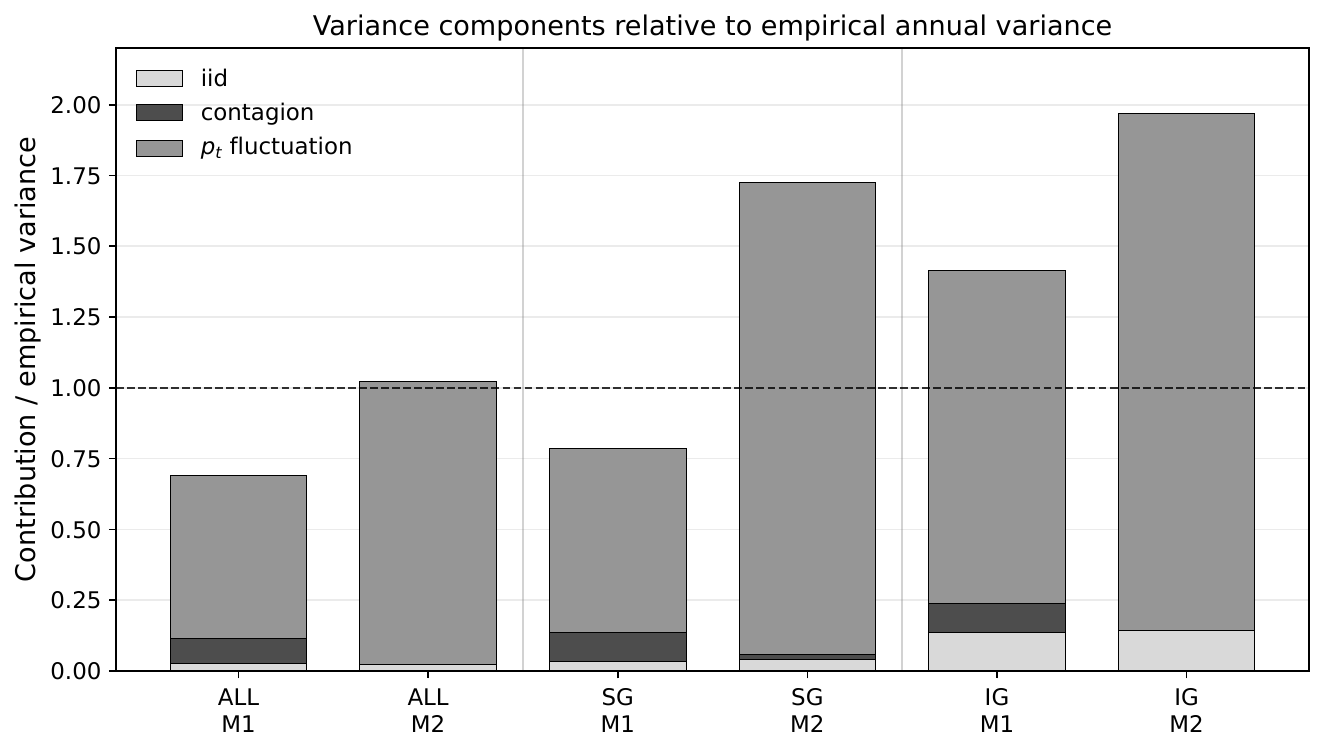}
\caption{
Visual representation of the full-sample variance decomposition in
Table~\ref{tab:variance_decomp_full}, normalized by the empirical variance of
the scaled annual default counts.
}
\label{fig:variance_decomp_full}
\end{figure}

For the IG class, the total model-implied variance exceeds the empirical scaled
variance under both extensions, reflecting the sensitivity of rare-default
regimes to small changes in $p_t$. Appendix~\ref{app:sub} examines the
stability of the same decomposition across subperiods.

\subsection{KL Projection with Environmental Fluctuations}
\label{subsec:environmental-kl}

We apply the KL-projection method introduced in
Sec.~\ref{subsec:kl-iid} to {infectious default-count distributions
with environmental fluctuations}. The purpose is to examine whether the
distributional signatures of contagion remain distinguishable from smooth
environmental mixtures once cross-period variation in $p_t$ is introduced.

We construct {environmentally mixed versions} of the three
representative Torri models introduced in Sec.~\ref{subsec:kl-iid},
corresponding to high, medium, and low values of $\pi_n$, as well as the
{Davis--Lo model with environmental fluctuations}. For each
infectious target distribution, we compute its KL projection onto the Vasicek
family and examine how the KL {divergence} changes as the relative
importance of environmental fluctuations increases.

{
For each target model, we summarize the relative importance of environmental
fluctuations by the variance contribution ratio
\begin{equation}
r
=
\frac{R_{p_t}}
{
R_{\mathrm{iid}}
+
R_{\mathrm{infect}}
+
R_{p_t}
}.
\label{eq:variance_ratio_r}
\end{equation}
Here, $r$ is defined for the target infectious model with environmental
fluctuations; it is not a parameter or variance decomposition of the Vasicek
projection family.
}

A decreasing KL {divergence} indicates that variation in $p_t$
absorbs the distributional signature of contagion. In contrast, a positive KL
gap indicates residual distributional structure that is not reproduced by the
smooth Vasicek-type environmental mixture.

\begin{figure}[t]
\centering
\includegraphics[width=0.72\linewidth]{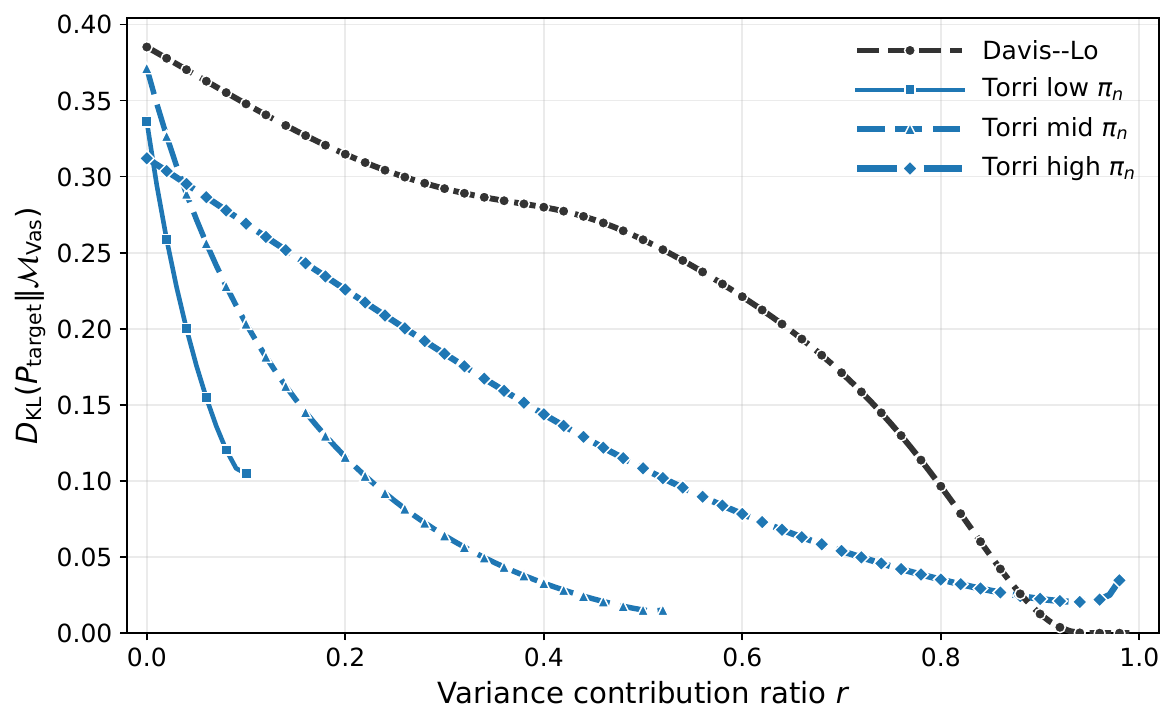}
\caption{
{
KL divergence from infectious target distributions with environmental
fluctuations to the best-fitting Vasicek family as a function of the variance
contribution ratio $r$.
}
The vertical axis reports
$D_{\mathrm{KL}}(P_{\mathrm{target}}\|\mathcal{M}_{\mathrm{Vas}})$,
where $P_{\mathrm{target}}$ is the {Davis--Lo or Torri target
distribution with environmental fluctuations} and
$\mathcal{M}_{\mathrm{Vas}}$ denotes the Vasicek family optimized by KL
projection.
}
\label{fig:environmental_kl_vasicek}
\end{figure}

Figure~\ref{fig:environmental_kl_vasicek} shows that the KL
{divergence} generally decreases as $r$ increases, indicating that
cross-period variation in $p_t$ absorbs part of the distributional signature
generated by contagion.
{
The low- and intermediate-$\pi_n$ Torri targets lose separation particularly
rapidly. 
The Davis--Lo target retains a larger KL gap over an intermediate
range of $r$, but this gap also decreases when $r$ becomes sufficiently large.
}

{
These calculations are illustrative comparisons of the distributional
relations among the model families. They are not class-specific empirical
calibrations and are not used as an additional model-selection criterion.
}

\subsection{Distributional Separation among the Fitted M0, M1, and M2 Models}
\label{subsec:fitted-kl}

{
The preceding KL-projection analysis characterizes, in a controlled setting,
how environmental fluctuations can reduce the distributional separation
between contagion-based models and a smooth environmental mixture.
We now examine the corresponding separation among the models fitted to the
empirical data.

For each credit class, the fitted parameters of M0, M1, and M2 are held fixed
at their best-found values. Unlike the representative fixed-portfolio
comparisons used in the distributional plots, the KL divergences are evaluated
at the actual portfolio size $n_t$ observed in each year. For models
$a,b\in\{0,1,2\}$, we define
\begin{equation}
D_{a\Vert b}
=
\frac{1}{T}
\sum_{t=1}^{T}
\sum_{h=0}^{n_t}
P_a(h\mid n_t)
\log
\frac{P_a(h\mid n_t)}
     {P_b(h\mid n_t)},
\end{equation}
where $P_a(h\mid n_t)$ denotes the fitted default-count distribution under
model M$a$ for year $t$. Thus, the reported divergences measure the average
distributional separation under the actual sequence of portfolio sizes used
in the likelihood analysis. Because the KL divergence is asymmetric,
both $D_{a\Vert b}$ and $D_{b\Vert a}$ are reported.

\begin{table}[htbp]
\centering
\caption{{
Mean directed KL divergences among the fitted M0, M1, and M2
default-count distributions, averaged over the observed annual portfolio
sizes $n_t$. Rows denote the target distribution $P_a$ and columns denote
the reference distribution $P_b$.}
}
\label{tab:fitted_kl_m012}

\begin{tabular}{lccc}
\hline
\multicolumn{4}{c}{\textbf{(a) ALL}}\\
\hline
Target $P_a$ & M0 & M1 & M2\\
\hline
M0 & 0.0000 & 0.2435 & 0.0000\\
M1 & 0.1941 & 0.0000 & 0.1941\\
M2 & 0.0000 & 0.2435 & 0.0000\\
\hline
\\[-1.5ex]

\hline
\multicolumn{4}{c}{\textbf{(b) SG}}\\
\hline
Target $P_a$ & M0 & M1 & M2\\
\hline
M0 & 0.0000 & 0.2585 & 0.1693\\
M1 & 0.2083 & 0.0000 & 0.2251\\
M2 & 0.1238 & 0.3501 & 0.0000\\
\hline
\\[-1.5ex]

\hline
\multicolumn{4}{c}{\textbf{(c) IG}}\\
\hline
Target $P_a$ & M0 & M1 & M2\\
\hline
M0 & 0.0000 & 0.0219 & 0.0000\\
M1 & 0.0189 & 0.0000 & 0.0189\\
M2 & 0.0000 & 0.0219 & 0.0000\\
\hline
\end{tabular}
\end{table}

The results closely mirror the likelihood-based identification pattern in
Sec.~\ref{sec:identification}. For ALL, M2 lies on the no-contagion boundary
and is therefore distributionally identical to M0, while M1 remains clearly
separated from both. For SG, all three fitted distributions are distinct:
both contagion extensions are separated from M0, and M1 and M2 also remain
substantially separated from each other. For IG, M2 again coincides with M0,
while the KL separation between M1 and M0 is much smaller than in ALL or SG,
consistent with the weak likelihood improvement of M1 in this class.

These results provide a distributional interpretation of the model-comparison
results. In ALL, the Davis--Lo extension generates a residual count-distribution
signature beyond the environmental null, whereas the Torri extension is fully
absorbed by the null. In SG, both contagion mechanisms generate distributions
that differ from M0, although M1 provides the substantially better empirical
fit. In IG, the fitted distributions are only weakly separated.

The KL divergences in Table~\ref{tab:fitted_kl_m012} are used only as a
descriptive measure of separation among the fitted count distributions.
They are not used as an additional model-selection criterion.}

\newpage

\section{Parameter Estimates for the Two-Component Gaussian-Mixture Environmental Null}
\label{app:m3_parameters}
{

The fitted M3 component parameters are reported in
Table~\ref{tab:m3_parameters}. They should be interpreted cautiously. The
mixture likelihood has nearby alternative modes for ALL and SG, while the IG
solution contains a nearly degenerate second component. Accordingly, M3 is used
here as a flexible environmental null for a robustness test, not as evidence for
two distinct economic regimes.}

\begin{table}[t]
\centering
\caption{{
Parameters of the cross-language validated best-found M3 candidates. The ordering
$\mu_1<\mu_2$ is imposed only for reporting. The IG value of $\sigma_2$ is
near zero, indicating a nearly degenerate component.}
}
\label{tab:m3_parameters}
\begin{tabular}{lrrrrr}
\hline
Class & $\mu_1$ & $\sigma_1$ & $\mu_2$ & $\sigma_2$ & $\pi$ \\
\hline
ALL & -2.580053 & 0.697083 & -2.411763 & 0.376634 & 0.199837 \\
SG  & -2.837232 & 0.895836 & -1.962688 & 0.337446 & 0.161096 \\
IG  & -3.419428 & 0.499547 & -2.608762 & $5.14\times10^{-8}$ & 0.883577 \\
\hline
\end{tabular}
\end{table}

\newpage

\section{Subperiod estimation and variance decomposition}
\label{app:sub}

Section~\ref{sec:data} documented differences in summary statistics and time-series
behavior across the two subperiods, 1950--1979 and 1980--2023.
We now examine whether the corresponding model-based interpretations
are stable across periods.

\subsection{Likelihood comparison by subperiod}

\begin{table}[htbp]
\centering
\caption{Maximum likelihood estimates in the static setting by subperiod.
Reported values are negative log-likelihoods (NLL) and selected parameters.
For each class and period, the best fit (lowest NLL) is highlighted in bold.}
\begin{tabular}{llcccc}
\hline
Model & Class (Period) & nll & $p$ & $q$ / $\rho_A$ & Remarks \\
\hline
Davis--Lo & ALL (1950--1979) & 63.57 & 0.00129 & $q=0.00132$ &  \\
Torri     & ALL (1950--1979) & 90.01 & 0.00295 & $u\approx0,\ v=0$ & boundary \\
Vasicek   & ALL (1950--1979) & \textbf{58.24} & 0.00255 & $\rho_A=0.0982$ &  \\

Davis--Lo & IG (1950--1979)  & \textbf{11.81} & 0.00017 & $q=0.00072$ &  \\
Torri     & IG (1950--1979)  & 13.96 & 0.00027 & $u=0.43,\ v=0$ & boundary \\
Vasicek   & IG (1950--1979)  & 12.91 & 0.00026 & $\rho_A=0.168$ &  \\

Davis--Lo & SG (1950--1979)  & 64.14 & 0.00445 & $q=0.00494$ &  \\
Torri     & SG (1950--1979)  & \textbf{57.27} & 0.00680 & $u=0.923,\ v=0.0209$ &  \\
Vasicek   & SG (1950--1979)  & 58.85 & 0.00905 & $\rho_A=0.144$ &  \\
\hline
Davis--Lo & ALL (1980--2023) & 223.13 & 0.00088 & $q=0.00391$ &  \\
Torri     & ALL (1980--2023) & 322.46 & 0.00961 & $u=0.984,\ v=0.0125$ &  \\
Vasicek   & ALL (1980--2023) &
{\textbf{214.61}} &
{0.01568} &
{$\rho_A=0.0548$} &  \\

Davis--Lo & IG (1980--2023)  & 87.85 & 0.00033 & $q=0.00067$ &  \\
Torri     & IG (1980--2023)  & 143.77 & 0.00098 & $u=0,\ v=0$ & boundary \\
Vasicek   & IG (1980--2023)  &
\textbf{84.71} &
0.00097 &
{$\rho_A=0.165$} &  \\

Davis--Lo & SG (1980--2023)  & 238.33 & 0.00148 & $q=0.01591$ &  \\
Torri     & SG (1980--2023)  & 323.16 & 0.02716 & $u=0.959,\ v=0.00926$ &  \\
Vasicek   & SG (1980--2023)  &
{\textbf{208.75}} &
{0.04188} &
{$\rho_A=0.0635$} &  \\
\hline
\end{tabular}
\label{tab:mle_results_subperiod}
\end{table}

\begin{table}[htbp]
\centering
\caption{Comparison of negative log-likelihoods (NLL) by subperiod. Lower values indicate better fit.}
\label{tab:nll_compare_subperiod}
\begin{tabular}{lccccc}
\hline
Class (Period) & Torri & Davis--Lo & Vasicek & M2 & M1 \\
\hline
ALL (1950--1979) & 90.01 & 63.57 & 58.24 & 58.24 & \textbf{58.16} \\
SG  (1950--1979) & 57.27 & 64.14 & 58.85 & \textbf{55.96} & 58.84 \\
IG  (1950--1979) & 13.96 & 11.81 & 12.91 & \textbf{11.72} & 11.81 \\
\hline
ALL (1980--2023) & 322.46 & 223.13 & {214.61} & 214.61 & \textbf{212.73} \\
SG  (1980--2023) & 323.16 & 238.33 & {\textbf{208.75}} & \textbf{208.75} & \textbf{208.75} \\
IG  (1980--2023) & 143.77 & 87.85 & 84.71 & 84.71 & \textbf{83.42} \\
\hline
\end{tabular}
\end{table}

Tables~\ref{tab:mle_results_subperiod} and
\ref{tab:nll_compare_subperiod} report the maximum-likelihood estimates
and the corresponding negative log-likelihood (NLL) values for each subperiod.
Table~\ref{tab:nll_compare_subperiod} additionally includes the environmental
extensions M1 and M2, whereas Table~\ref{tab:mle_results_subperiod} reports the three
static one-period models.

The relative ranking of models varies across classes and periods.
In particular, the Torri specification attains the best likelihood for the SG class in the early period (1950--1979),
but this advantage is not uniform across classes and disappears for ALL and IG.
This already suggests that likelihood alone may not be sufficient
to assess whether the underlying mechanism is consistently identified
from aggregated data.

\subsection{Variance decomposition by subperiod}

\begin{figure}[htbp]
\centering
\includegraphics[width=\textwidth]{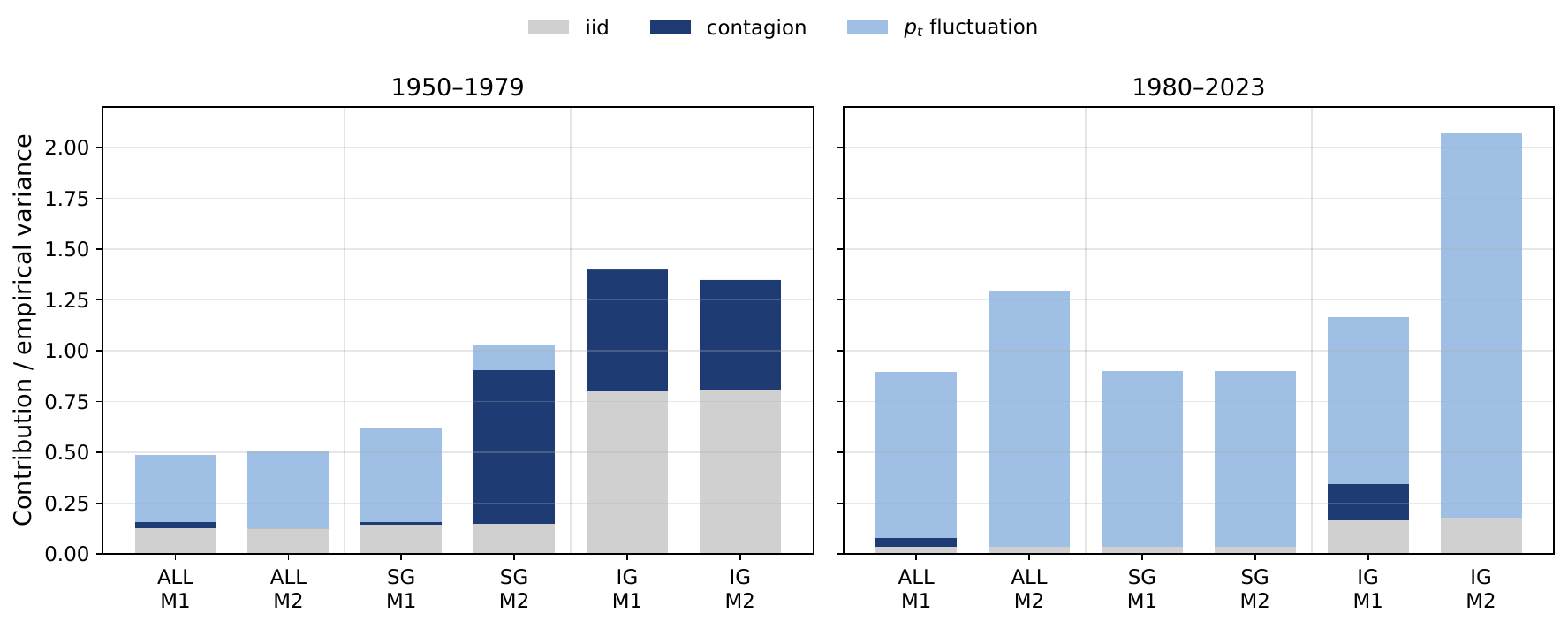}
\caption{Variance decomposition by subperiod, 1950--1979 and 1980--2023.
Contributions are normalized by the empirical variance.}
\label{fig:var_decomp_subperiods}
\end{figure}

As a supplementary temporal robustness check, Figure~\ref{fig:var_decomp_subperiods} decomposes the empirical variance into contributions from binomial variation, contagion, and time variation in the default probability $p_t$ for the two subperiods.

A clear structural difference emerges across periods.
In the early period (1950--1979), the contagion component plays a major
role in several cases. In particular, for the SG class under M2,
the infection term accounts for a large normalized contribution in the early period.

However, this dominance is not consistent across classes.
In the same period, the ALL results do not exhibit a comparable
contribution, despite ALL being the aggregate of SG and IG.
This discrepancy suggests that M2 is less stable at the aggregate level.

By contrast, M1 yields more coherent decompositions
across classes, with a positive infection contribution retained more consistently
across ALL, SG, and IG.

In the later period (1980--2023), the contribution from time variation
in $p_t$ becomes dominant across most specifications, while the contagion
component is substantially reduced.
M2 largely attributes the variance to $p_t$ fluctuations,
whereas M1 still retains a non-negligible contagion
component, particularly in the IG class.

This comparison is consistent with the main conclusion: M1 yields a
more stable decomposition across classes and periods, whereas M2 is
more sensitive to the sample period.

\section*{Data and code availability}

The data analysis and simulation codes used in this study,
including all scripts used to generate the figures and tables,
are available in a public GitHub repository:

\begin{center}
\url{https://github.com/shintaromori/contagion-vs-macro-defaults}
\end{center}

The empirical default data analyzed in this paper are not included in the
repository, as they are derived from proprietary historical default datasets
and therefore cannot be publicly shared. The repository provides the full
analysis and simulation code so that the results can be reproduced by
researchers with access to comparable data sources.

\bibliographystyle{apsrev4-2}
\bibliography{references}

@book{MantegnaStanley1999,
  author    = {Mantegna, Rosario N. and Stanley, H. Eugene},
  title     = {An Introduction to Econophysics: Correlations and Complexity in Finance},
  publisher = {Cambridge University Press},
  year      = {1999},
  doi       = {10.1017/CBO9780511755767},
  isbn      = {978-0511755767}
}

@article{Galam2008,
  author  = {Galam, Serge},
  title   = {Sociophysics: A Review of Galam Models},
  journal = {Int. J. Mod. Phys. C},
  volume  = {19},
  number  = {3},
  pages   = {409--440},
  year    = {2008},
  doi     = {10.1142/S0129183108012297}
}

@article{Lux1995,
  author  = {Lux, Thomas},
  title   = {Herd Behaviour, Bubbles and Crashes},
  journal = {Econ. J.},
  volume  = {105},
  number  = {431},
  pages   = {881--896},
  year    = {1995},
  doi     = {10.2307/2235156}
}

@article{LuxMarchesi1999,
  author  = {Lux, Thomas and Marchesi, Michele},
  title   = {Scaling and Criticality in a Stochastic Multi-Agent Model of a Financial Market},
  journal = {Nature},
  volume  = {397},
  pages   = {498--500},
  year    = {1999},
  doi     = {10.1038/17290}
}

@article{Alfarano2005,
  author  = {Alfarano, Simone and Lux, Thomas and Wagner, Friedrich},
  title   = {Estimation of Agent-Based Models: The Case of an Asymmetric Herding Model},
  journal = {Comput. Econ.},
  volume  = {26},
  number  = {1},
  pages   = {19--49},
  year    = {2005},
  month   = {aug},
  doi     = {10.1007/s10614-005-6415-1}
}

@article{Bouchaud2002,
  author  = {Bouchaud, Jean-Philippe and M{\'e}zard, Marc and Potters, Marc},
  title   = {Statistical Properties of Stock Order Books: Empirical Results and Models},
  journal = {Quant. Finance},
  volume  = {2},
  number  = {4},
  pages   = {251--256},
  year    = {2002},
  doi     = {10.1088/1469-7688/2/4/301}
}

@article{FernandezGracia2014,
  author  = {Fernandez-Gracia, J. and Suchecki, K. and Ramasco, J. J. and SanMiguel, M. and Egu{\'i}luz, V. M.},
  title   = {Is the Voter Model a Model for Voters?},
  journal = {Phys. Rev. Lett.},
  volume  = {112},
  pages   = {158701},
  year    = {2014},
  doi     = {10.1103/PhysRevLett.112.158701}
}

@article{MoriHisakadoTakahashi2012,
  author  = {Mori, Shintaro and Hisakado, Masato and Takahashi, Taiki},
  title   = {Phase Transition to a Two-Peak Phase in an Information-Cascade Voting Experiment},
  journal = {Phys. Rev. E},
  volume  = {86},
  pages   = {026109},
  year    = {2012},
  doi     = {10.1103/PhysRevE.86.026109}
}

@article{SmolyakHavlin2022,
  author  = {Smolyak, Alex and Havlin, Shlomo},
  title   = {Three Decades in Econophysics---From Microscopic Modelling to Macroscopic Complexity and Back},
  journal = {Entropy},
  year    = {2022},
  volume  = {24},
  number  = {2},
  pages   = {271},
  doi     = {10.3390/e24020271}
}

@book{Schonbucher2003,
  author    = {Sch{\"o}nbucher, Philipp J.},
  title     = {Credit Derivatives Pricing Models: Models, Pricing and Implementation},
  publisher = {John Wiley \& Sons},
  year      = {2003}
}

@article{DavisLo2001,
  author  = {Davis, Mark H. A. and Lo, Violet},
  title   = {Infectious defaults},
  journal = {Quant. Finance},
  volume  = {1},
  number  = {4},
  pages   = {382--387},
  year    = {2001},
  doi     = {10.1080/713665832}
}

@article{Vasicek1991,
  author  = {Vasicek, Oldrich A.},
  title   = {Limiting Loan Loss Probability Distribution},
  journal = {KMV Corporation},
  year    = {1991},
  note    = {Working paper}
}

@article{Vasicek2002,
  author  = {Vasicek, Oldrich A.},
  title   = {Loan portfolio value},
  journal = {Risk},
  volume  = {15},
  number  = {12},
  pages   = {160--162},
  year    = {2002}
}

@article{DasDuffieKapadiaSaita2007,
  author  = {Das, Sanjiv R. and Duffie, Darrell and Kapadia, Nikunj and Saita, Leandro},
  title   = {Common Failings: How Corporate Defaults Are Correlated},
  journal = {J. Finance},
  year    = {2007},
  volume  = {62},
  number  = {1},
  pages   = {93--117},
  doi     = {10.1111/j.1540-6261.2007.01202.x}
}

@article{DuffieEcknerHorelSaita2009,
  author  = {Duffie, Darrell and Eckner, Andreas and Horel, Guillaume and Saita, Leandro},
  title   = {Frailty Correlated Default},
  journal = {J. Finance},
  volume  = {64},
  number  = {5},
  pages   = {2089--2123},
  year    = {2009},
  doi     = {10.1111/j.1540-6261.2009.01485.x}
}

@article{AzizpourGieseckeSchwenkler2018,
  author  = {Azizpour, S. and Giesecke, K. and Schwenkler, G.},
  title   = {Exploring the Sources of Default Clustering},
  journal = {J. Financ. Econ.},
  volume  = {129},
  number  = {1},
  pages   = {154--183},
  year    = {2018},
  doi     = {10.1016/j.jfineco.2018.04.008}
}

@article{HisakadoKitsukawaMori2006,
  author  = {Hisakado, Masato and Kitsukawa, Katsuhiro and Mori, Shintaro},
  title   = {Correlated Binomial Models of Credit Risk},
  journal = {Journal of Physics A: Mathematical and General},
  volume  = {39},
  number  = {50},
  pages   = {15365--15378},
  year    = {2006},
  doi     = {10.1088/0305-4470/39/50/005}
}

@article{Mori2008,
  author  = {Mori, Shintaro and Kitsukawa, Kenji and Hisakado, Masato},
  title   = {Correlation Structures of Correlated Binomial Models and Implied Default Distribution},
  journal = {Journal of the Physical Society of Japan},
  volume  = {77},
  number  = {11},
  pages   = {114802},
  year    = {2008},
  doi     = {10.1143/JPSJ.77.114802}
}

@book{DeFinetti1974,
  author    = {de Finetti, Bruno},
  title     = {Theory of Probability},
  publisher = {Wiley},
  year      = {1974}
}

@article{TorriGiacomettiFarina2026,
title = {Modeling portfolio loss distribution under infectious defaults and immunization},
journal = {Commun. Nonlinear Sci. Numer. Simul.},
volume = {159},
pages = {109886},
year = {2026},
issn = {1007-5704},
doi = {https://doi.org/10.1016/j.cnsns.2026.109886},
url = {https://www.sciencedirect.com/science/article/pii/S1007570426002467},
author = {Gabriele Torri and Rosella Giacometti and Gianluca Farina}
}

@article{Hawkes1971,
  author  = {Hawkes, A. G.},
  title   = {Spectra of Some Self-Exciting and Mutually Exciting Point Processes},
  journal = {Biometrika},
  volume  = {58},
  number  = {1},
  pages   = {83--90},
  year    = {1971},
  doi     = {10.1093/biomet/58.1.83}
}

@article{Kirchner2017,
  author  = {Kirchner, Matthias},
  title   = {An estimation procedure for the Hawkes process},
  journal = {Quant. Finance},
  volume  = {17},
  number  = {4},
  pages   = {571--595},
  year    = {2017},
  doi     = {10.1080/14697688.2016.1211312}
}

@article{BlancDonierBouchaud2017,
  author  = {Blanc, Pierre and Donier, Jonathan and Bouchaud, Jean-Philippe},
  title   = {Quadratic Hawkes processes for financial prices},
  journal = {Quant. Finance},
  volume  = {17},
  number  = {2},
  pages   = {171--188},
  year    = {2017},
  doi     = {10.1080/14697688.2016.1193215}
}

@article{ErraisGieseckeGoldberg2010,
  author  = {Errais, Eymen and Giesecke, Kay and Goldberg, Lisa R.},
  title   = {Affine Point Processes and Portfolio Credit Risk},
  journal = {SIAM J. Financial Math.},
  volume  = {1},
  number  = {1},
  pages   = {642--665},
  year    = {2010},
  doi     = {10.1137/090771272}
}

@misc{Mori2026TemporalCoarseGraining,
  author        = {Mori, Shintaro},
  title         = {Temporal Coarse-Graining of Latent Default-Probability Paths Generates Effective Default Correlation},
  year          = {2026},
  eprint        = {2606.12446},
  archivePrefix = {arXiv},
  primaryClass  = {q-fin.ST},
  doi           = {10.48550/arXiv.2606.12446},
  note          = {arXiv preprint arXiv:2606.12446}
}

@article{KoyamaShinomoto2020,
  author  = {Shinsuke Koyama and Shigeru Shinomoto},
  title   = {Statistical physics of discovering exogenous and endogenous factors in a chain of events},
  journal = {Physical Review Research},
  volume  = {2},
  pages   = {043358},
  year    = {2020},
  doi     = {10.1103/PhysRevResearch.2.043358}
}

@article{KanazawaSornette2020,
  author  = {Kiyoshi Kanazawa and Didier Sornette},
  title   = {Field master equation theory of the self-excited Hawkes process},
  journal = {Physical Review Research},
  volume  = {2},
  pages   = {033442},
  year    = {2020},
  doi     = {10.1103/PhysRevResearch.2.033442}
}

@article{KullbackLeibler1951,
  author  = {Kullback, Solomon and Leibler, Richard A.},
  title   = {On Information and Sufficiency},
  journal = {Ann. Math. Statist.},
  volume  = {22},
  number  = {1},
  pages   = {79--86},
  year    = {1951},
  doi     = {10.1214/aoms/1177729694}
}

@book{CoverThomas2006,
  author    = {Cover, Thomas M. and Thomas, Joy A.},
  title     = {Elements of Information Theory},
  edition   = {2},
  publisher = {Wiley-Interscience},
  address   = {Hoboken, NJ},
  year      = {2006}
}

@book{Vasicek2015,
  author    = {Vasicek, Oldrich A.},
  title     = {Finance, Economics and Mathematics},
  publisher = {John Wiley \& Sons},
  year      = {2015},
  isbn      = {9781119186205}
}

@article{GotoTakayasuTakayasu2017,
  author  = {Hayato Goto and Hideki Takayasu and Misako Takayasu},
  title   = {Estimating risk propagation between interacting firms on inter-firm complex network},
  journal = {PLoS ONE},
  volume  = {12},
  number  = {10},
  pages   = {e0185712},
  year    = {2017},
  doi     = {10.1371/journal.pone.0185712}
}

@inproceedings{SoumaFujiwaraAoyama2005,
  author    = {Wataru Souma and Yoshi Fujiwara and Hideaki Aoyama},
  title     = {Shareholding Networks in Japan},
  booktitle = {AIP Conference Proceedings},
  volume    = {776},
  pages     = {298--307},
  year      = {2005},
  doi       = {10.1063/1.1985396}
}

\end{document}